\documentclass[
aps,
pra,
nosuperscriptaddress,
twocolumn,
notitlepage
]{revtex4-2}
\usepackage{float}
\usepackage{graphicx}
\usepackage[svgnames,psnames]{xcolor}
\usepackage[colorlinks,
citecolor=FireBrick,
linkcolor=Blue,
linktocpage,
unicode]{hyperref}

\usepackage{amsthm,amsmath,amssymb,amsbsy}
\usepackage{bbm}
\usepackage{physics}

\newcommand{\mc}[1]{\mathcal{#1}}
\newcommand{\ox}{\otimes}
\newcommand{\id}{\mathbbm{1}}
\newcommand{\variance}{\mathrm{Var}}
\newcommand{\gibbs}{\rho_\mathrm{th}}
\newcommand{\bures}{\theta_B}

\newcommand{\inlineheading}[1]{\textit{#1}.---\ignorespaces}

\newtheorem{lem}{Lemma}

\newtheorem{res}{Result}

\begin{document} 

\title{Quantum speed limit for states and observables of perturbed open systems}

\author{Benjamin Yadin}
\author{Satoya Imai}
\author{Otfried G\"uhne}

\affiliation{Naturwissenschaftlich-Technische Fakult\"at, Universit\"at Siegen, Walter-Flex-Stra\ss e 3, 57068 Siegen, Germany}
    
\begin{abstract}
    Quantum speed limits provide upper bounds on the rate with which a quantum system can move away from its initial state.
    Here, we provide a different kind of speed limit, describing the divergence of a perturbed open system from its unperturbed trajectory.
    In the case of weak coupling, we show that the divergence speed is bounded by the quantum Fisher information under a perturbing Hamiltonian, up to an error which can be estimated from system and bath timescales.
    We give two applications of our speed limit.
    Firstly, it enables experimental estimation of quantum Fisher information in the presence of decoherence that is not fully characterised.
    Secondly, it implies that large quantum work fluctuations are necessary for a thermal system to be driven quickly out of equilibrium under a quench.
    Moreover, it can be used to bound the response to perturbations of expectation values of observables in open systems.
\end{abstract}

\date{\today}
\maketitle

The quantum time-energy uncertainty relation was first put on a rigorous footing by Mandelstam and Tamm~\cite{Mandelstam1945Uncertainty}, who showed that the time for a quantum state to evolve to an orthogonal one is limited according to its energy uncertainty.
Since then, many generalisations and variations have been derived, including the Margolus-Levitin bound~\cite{Margolus1998Maximum} involving mean energy.
Such inequalities are referred to as quantum speed limits.
We now have extensions to mixed states and driven and open systems~\cite{Jones2010Geometric,delCampo2013Quantum,Pires2016Generalized,Deffner2013Quantum,Marvian2016Quantum}, and an understanding of the connection to the geometry of quantum state spaces~\cite{Uhlmann1992Energy,Braunstein1994Statistical,Frowis2012Kind}.

Quantum speed limits imply bounds on information processing rates~\cite{Bremermann1965Quantum,Lloyd2000Ultimate,Deffner2020Quantum} and maximum physically allowable rates of communication~\cite{Bekenstein1974Generalized}.
There are also applications within quantum thermodynamics, including bounding entropy production rates~\cite{Deffner2010Generalized}, heat engine efficiency and power \cite{Abah2017Energy,Mukherjee2016Speed}, and battery charging rates~\cite{Binder2015Quantacell,Campaioli2017Enhancing,Mohan2022Quantum2,Julia-Farre2020Bounds}.
There is an intimate relation between speed limits and metrology, in which the best precision in estimating parameter encoded into a state depends on how quickly the state evolves~\cite{Gessner2018Statistical,Zhang2017Detecting}.

A central quantity in metrology, the quantum Fisher information (QFI)~\cite{Braunstein1994Statistical,Paris2009Quantum}, is interpretable as a (squared) speed in state space, and can be used to quantify many important properties of quantum states.
Sufficiently large QFI demonstrates many-body entanglement~\cite{Amico2008Entanglement,Hyllus2012Fisher,Toth2012Multipartite,Morris2020Entanglement,Pezze2009Entanglement,Pezze2015Witnessing} and steering~\cite{Yadin2021Metrological}; similarly, QFI can be used as a measure of coherence in a given basis~\cite{Yadin2016General,Marvian2020Coherence,Marvian2022Operational}, of macroscopic quantumness~\cite{Frowis2018Macroscopic} and of optical nonclassicality~\cite{Yadin2018Operational,Kwon2019Nonclassicality}, and can witness general quantum resources~\cite{Tan2021Fisher}.
Therefore, it is desirable to experimentally measure lower bounds to QFI.
One common method, among others~\cite{Girolami2014Observable,Rath2021Quantum} is to adopt speed limits to estimate the QFI from a measure of distance between an initial state and one evolved for a short time~\cite{Strobel2014Fisher,Frowis2016Lower}.

In addition, it can be useful to bound the speed at which two quantum states separate when they undergo different dynamics.
Examples of applications include the discrimination of unitary operations~\cite{Becker2021Energy}, the performance of adiabatic quantum computation~\cite{Suzuki2020Performance}, fidelity of quantum control~\cite{Hatomura2022Performance}, dynamics of entanglement~\cite{Pandey2023Fundamental}, and multi-parameter metrology~\cite{Albarelli2022Probe}.

In this work, we devise a novel type of speed limit that describes the response of a Markovian open quantum system to a perturbation to its dynamics.
The inequality upper-bounds the distance between the perturbed and unperturbed trajectories in state space in terms of the QFI of the system with respect to the perturbation.
Importantly, this holds under minimal assumptions without detailed knowledge of the dynamics.
For a system weakly coupled to its environment, the speed limit is given in terms of the QFI under a perturbing Hamiltonian, up to an error bounded in terms of relevant physical timescales.
We show how this may be used for an experimental lower bound on the QFI.
We then provide an application to the thermodynamics of systems perturbed out of equilibrium, showing that quantum fluctuations in the work performed during a sudden quench are required for fast departure from the initial state.
Finally, we give an application to linear response by bounding changes in expectation values under perturbations.

\inlineheading{Preliminaries}
The Mandelstam-Tamm bound~\cite{Mandelstam1945Uncertainty} relates the energy variance $\variance(\psi,H) = \ev{H^2}{\psi} - \ev{H}{\psi}^2$ of a pure state $ \ket{\psi}$ to the time $\tau$ it needs to evolve to an orthogonal one under Hamiltonian $H$:
\begin{equation} \label{eqn:mt_bound}
    \tau \geq \frac{\pi}{2 \sqrt{\variance(\psi,H)}}.
\end{equation}
(We work in units where $\hbar = 1$ throughout.)
Therefore a large energy variance is necessary to evolve quickly to an orthogonal state.
This result has since been strengthened to account for mixed states and non-orthogonality.
The Uhlmann bound~\cite{Uhlmann1992Energy} involves the fidelity $F(\rho_0, \rho_t) := \tr \sqrt{\sqrt{\rho_0} \rho_t\sqrt{\rho_0}}$ between the initial and final states, $\rho_0$ and $\rho_t = e^{-itH} \rho_0 e^{itH}$, recast into the Bures angle $\bures(\rho,\sigma) := \arccos F(\rho, \sigma)$.

Instead of energy variance, we require the QFI of the system.
Most generally, QFI measures the sensitivity of a continuously parameterised family of states to small changes in a parameter~\cite{Paris2009Quantum}.
Here, we consider the time parameter, so the QFI is a function of the state $\rho_t$ and its derivative $\frac{\dd \rho_t}{\dd t}$.
Throughout this paper, we consider evolutions generated by Gorini-Kossakowski-Sudarshan-Lindblad (GKSL) superoperators~\cite{Gorini1976Completely,Lindblad1979On} $\frac{\dd\rho_t}{\dd t} = \mc{L}_t(\rho_t)$, for which the QFI $\mc{F}(\rho_t,\mc{L}_t)$ is a function of $\rho_t$ and $\mc{L}_t$.
One definition is expressed in terms of the spectral decomposition ${\rho_t = \sum_i \lambda_i(t) \dyad{\psi_i(t)}}$: $\mc{F}(\rho_t, \mc{L}_t) = 2 \sum_{i,j:\; \lambda_i(t) + \lambda_j(t) >0} \frac{\abs{ \mel{\psi_i(t)}{\mc{L}_t(\rho_t)}{\psi_j(t)} }^2}{\lambda_i(t) + \lambda_j(t)}$.
The Uhlmann bound is
\begin{equation} \label{eqn:uhlmann_bound}
    \bures(\rho_0, \rho_t) \leq \frac{1}{2} \int_0^t \dd s \; \sqrt{\mc{F}(\rho_s,\mc{H}_s)} \quad \forall \, t \geq 0,
\end{equation}
where $\mc{H}_t(\cdot) = -i[H_t, \cdot]$ generates time evolution under the time-dependent Hamiltonian $H_t$.
This bound derives ultimately from the infinitesimal expansion of the Bures angle as a metric on state space, $\bures(\rho_t,\rho_{t+\dd t})^2 = \frac{1}{4} \mc{F}(\rho_t, \mc{L}_t) \dd t^2$, with the finite Bures angle being the length of a geodesic between two points~\cite{Braunstein1994Statistical}.
Eq.~\eqref{eqn:mt_bound} can be derived from Eq.~\eqref{eqn:uhlmann_bound}~\cite{Deffner2010Generalized}; one sees that the square-root QFI may be interpreted as a ``statistical speed"~\cite{Paris2009Quantum,Gessner2018Statistical}.

\inlineheading{Perturbation speed limit}
Here, we prove the main result for a system undergoing arbitrary Markovian dynamics with a perturbation.
We take the common definition equating Markovianity with divisibility, namely that the mapping $\mc{N}_{t_1,t_0}$ of states between any times $t_0<t_1$ is completely positive and trace-preserving, and satisfies $\mc{N}_{t_2,t_0} = \mc{N}_{t_2,t_1}\mc{N}_{t_1,t_0}$ for all $t_0\leq t_1 \leq t_2$.
This is equivalent to the dynamics being dictated by a GKSL generator $\mc{L}_t$~\cite{Rivas2014Quantum}.

\begin{figure}[t]
    \includegraphics[width=0.35\textwidth]{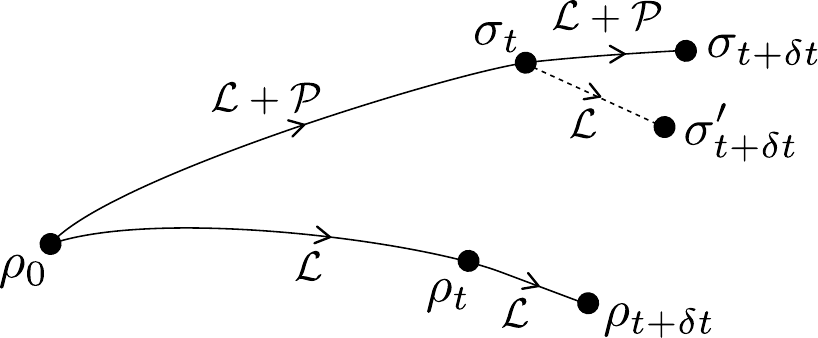}
    \caption{Illustration of the trajectories used in the proof of the main speed limit Eq.~\eqref{eqn:main_bound}.}
    \label{fig:trajectories}
\end{figure}

\begin{res} \label{res:main}
    Consider a system starting in state $\rho_0$ which may evolve along one of two trajectories: i) Markovian free evolution, $\frac{\dd \rho_t}{\dd t} = \mc{L}_t(\rho_t)$; or ii) perturbed evolution, $\frac{\dd \sigma_t}{\dd t} = \mc{L}'_t(\sigma_t) = \mc{L}_t(\sigma_t) + \mc{P}_t(\sigma_t)$ (satisfying the initial condition $\sigma_0 = \rho_0$).
    The Bures angle between the trajectories satisfies
    \begin{equation} \label{eqn:main_bound}
        \bures(\rho_t, \sigma_t) \leq \frac{1}{2} \int_0^t \dd s \; \sqrt{\mc{F}(\sigma_s, \mc{P}_s)} \quad \forall \, t \geq 0.
    \end{equation}
    \begin{proof}
        Here we summarise the proof detailed in Appendix~\ref{app:main_derivation}.
        We use three facts about the Bures angle: i) the triangle inequality, ii) contractivity under quantum channels~\cite{NielsenChuang_ch9}, and iii) its infinitesimal expansion, stated above.
        At time $t$, consider $\rho_t,\, \sigma_t$ and their corresponding time-evolved states $\rho_{t+\delta t},\,\sigma_{t+\delta t}$ a short time $\delta t$ later.
        In addition, consider instead evolving $\sigma_t$ under the \emph{unperturbed} dynamics for time $\delta t$, giving $\sigma'_{t+\delta t}$ (see Fig.~\ref{fig:trajectories}).
        To lowest order, $\rho_{t+\delta t} = \rho_t + \delta t \mc{L}_t(\rho_t) + \mc{O}(\delta t^2)$, $\sigma_{t+\delta t} = \sigma_t + \delta t \mc{L}_t(\sigma_t) + \delta t \mc{P}_t(\sigma_t) + \mc{O}(\delta t^2)$, and $\sigma'_{t+\delta t} = \sigma_t + \delta t \mc{L}_t(\sigma_t) + \mc{O}(\delta t^2)$.
        The triangle inequality gives
        \begin{equation} \label{eqn:triangle}
            \bures(\rho_{t+\delta t}, \sigma_{t+\delta t}) \leq \bures(\rho_{t+\delta t}, \sigma'_{t+\delta t}) + \bures(\sigma'_{t+\delta t}, \sigma_{t + \delta t}).
        \end{equation}
        For the first term on the right-hand side of Eq.~\eqref{eqn:triangle}, we use that $\rho_{t+\delta t}$ and $\sigma'_{t+\delta t}$ have been evolved for time $\delta t$ under the same dynamics comprising the channel $\mc{N}_{t+\delta t,t}$.
        Contractivity of $\bures$ therefore implies $\bures(\rho_{t+\delta t}, \sigma'_{t+\delta t})  = \bures(\mc{N}_{t+\delta t,t}(\rho_t), \mc{N}_{t+\delta t,t}(\sigma_t)) \leq \bures(\rho_t,\sigma_t).$
        For the second term, we use the infinitesimal form of $\bures$, and that $\sigma_{t+\delta t} - \sigma'_{t+\delta t} = \delta t \mc{P}_t(\sigma_t) + \mc{O}(\delta t^2)$, to write $\bures(\sigma'_{t+\delta t}, \sigma_{t+\delta t}) = \frac{\delta t}{2} \sqrt{\mc{F}(\sigma_{t+\delta t}, \mc{P}_t)} + \mc{O}(\delta t^2).$
        Putting these into Eq.~\eqref{eqn:triangle}, $\bures(\rho_{t+\delta t}, \sigma_{t+\delta t}) \leq \bures(\rho_t,\sigma_t) + \frac{\delta t}{2} \sqrt{\mc{F}(\sigma_{t+\delta t}, \mc{P}_t)} + \mc{O}(\delta t^2)$.
        Subtracting the first term on the right, dividing by $\delta t$ and taking $\delta t \to 0$ gives $\frac{\dd \bures(\rho_t, \sigma_t)}{\dd t} \leq \frac{1}{2} \sqrt{\mc{F}(\sigma_t, \mc{P}_t)}$; integrating gives the result.
    \end{proof}
\end{res}

In the case $\mc{L}_t = 0$, $\rho_t=\rho_0$ is stationary and the bound reduces to a previously known one~\cite{Taddei2013Quantum}.
The closed-system case is obtained with Hamiltonian dynamics $\mc{L}_t = \mc{H}_t$ and $\mc{P}_t(\cdot) = v\mc{V}_t(\cdot) := -i v[V_t, \cdot]$.
Note that the bound in this case is equivalent to Uhlmann's~\eqref{eqn:uhlmann_bound}, as can be seen by moving to the interaction picture.
Thus Eq.~\eqref{eqn:main_bound} generalises previous speed limits.
The relevant statistical speed measures sensitivity of the system to the perturbation.

\begin{figure}[t]
    \includegraphics[width=0.45\textwidth]{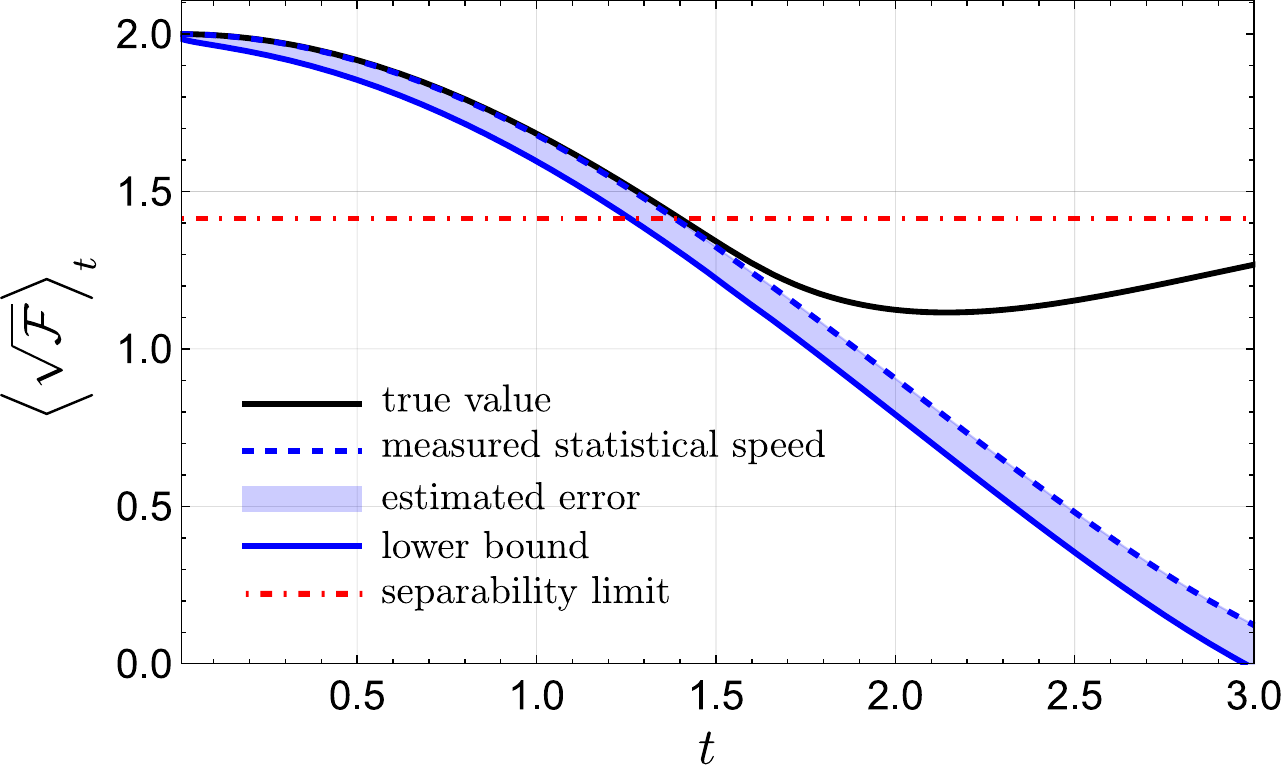}
        \caption{Two-qubit example with local dephasing noise, showing how the time-averaged value of $\sqrt{\mc{F}(\eta_s,\mc{V})}$ from $s=0$ to $t$ can be lower-bounded using the speed limit Eq.~\eqref{eqn:weak_coupling_speed}.
        The initial state $\frac{\ket{00}+\ket{11}}{\sqrt{2}}$ is maximally entangled.
        In units of $h=1$, we take $\lambda = \gamma = v = 0.1$ and $\epsilon \leq 4\lambda^2\gamma/h = 0.004$.
        The measured statistical speed is the left-hand side of Eq.~\eqref{eqn:witness}, taking a Bell-basis measurement $\{\frac{\ket{00} \pm \ket{11}}{\sqrt{2}}, \frac{\ket{01} \pm \ket{10}}{\sqrt{2}} \}$.
        The estimated error $\frac{2\Delta_\mathrm{est}(t)}{vt}$ (shaded area) is subtracted to give the lower bound.
        }
        \label{fig:qubits}
\end{figure}

Uhlmann's bound, relying on the triangle inequality, is saturated by geodesics~\cite{Uhlmann1992Energy}.
Eq.~\eqref{eqn:main_bound} additionally uses contractivity of $\bures$ under quantum channels, thus can only be saturated if $\mc{L}_t$ causes no contraction between the trajectories.
This will not generally hold; however, our later example in Fig.~\ref{fig:qubits} shows that the bound may be practically quite tight.

Using a similar method, we prove a related \emph{observable} speed limit.
This bounds the difference in expectation value of any observable $A$ between the trajectories, extending previous speed limits for observables~\cite{GarciaPintos2022Unifying,Mohan2022Quantum2} to the perturbation setting.
\begin{res} \label{res:obs_speed_lim}
    Using the same assumptions as Result~\ref{res:main}, for any observable $A$,
    \begin{align} \label{eqn:obs_lim}
        \abs{ \frac{\dd}{\dd t} \tr[A(\rho_t - \sigma_t)] } & \leq 2\norm*{\mc{L}_t^\dagger(A)} \, D_{\tr}(\rho_t,\sigma_t) \nonumber \\
            & \quad + \sqrt{\variance(\sigma_t, A) \mc{F}(\sigma_t, \mc{P}_t)},
    \end{align}
    where $\mc{L}_t^\dagger$ is the adjoint map of $\mc{L}_t$, $\norm*{\cdot}$ denotes the largest singular value, and $D_{\tr}(\rho,\sigma) = \frac{1}{2}\norm{\rho-\sigma}_1$ is the trace distance.
\end{res}
(See proof in Appendix~\ref{app:obs_speedlim}.)
A notable difference with Eq.~\eqref{eqn:main_bound} is the additional ``drift'' term depending on $\mc{L}_t$.

The power of Result~\ref{res:main} comes from requiring no detailed information about the unperturbed dynamics (in contrast to Result~\ref{res:obs_speed_lim}); such details only appear implicitly via the evolution of the perturbed trajectory.
For the remainder of this paper, we assume (typical in quantum control) that the perturbation comes from a controlled change to the system's Hamiltonian, $H_t \to H_t + v V_t$ (including the constant $v$ to quantify the size of the perturbation).
For many applications (see later sections), one is interested in QFI with respect to $V_t$; however, the resulting perturbation to the master equation $\mc{P}_t$ could contain additional terms.
The identification of $\mc{P}_t$ with $v \mc{V}_t$ may be justified in the singular coupling limit~\cite{lidar2019lecture} and in collision models of open system dynamics~\cite{Ciccarello2022Quantum}.
However, in weak coupling, a change to the system's Hamiltonian generally adds an additional perturbation to the master equation.
We therefore now study the error incurred by the approximation $\mc{P}_t \approx v \mc{V}_t$, and correspondingly the use of QFI with respect to $V_t$ in the right-hand side of Eq.~\eqref{eqn:main_bound}.
We consider time-independent dynamics for simplicity.\\

\inlineheading{Weak coupling}
To address this, we now consider a system weakly coupled to a Markovian environment and derive the error incurred by approximating the true perturbed trajectory with one where the dissipative part of the dynamics is unchanged.
Such situations are ubiquitious in experiments encompassing discrete-\cite{Leibfried2003Quantum} and continuous-variable~\cite{Rivas2010Markovian} systems.
We assume a standard weak-coupling master equation with secular approximation~\cite{Breuer2007Theory}, $\frac{\dd \rho_t}{\dd t} = \mc{L}(\rho_t) = -i[H + H_\mathrm{LS}, \rho_t] + \mc{D}(\rho_t)$, where the Lamb shift Hamiltonian $H_\mathrm{LS}$ and dissipator $\mc{D}$ are given by
$H_\mathrm{LS} = \lambda^2 \sum_{\omega,\alpha,\beta} S_{\alpha \beta}(\omega) A_\alpha^\dagger(\omega) A_\beta(\omega)$ and
$\mc{D}(\rho) = \lambda ^2 \sum_{\omega, \alpha,\beta} \gamma_{\alpha \beta}(\omega) \Big[  A_\beta(\omega) \rho A_\alpha^\dagger(\omega) - \frac{1}{2}\left\{ A_\alpha^\dagger(\omega) A_\beta(\omega), \rho \right\} \Big]$,
involving real and imaginary parts $\gamma_{\alpha \beta},\, S_{\alpha \beta}$ of the bath correlation function and jump operators $A_\alpha(\omega)$ assocated with Bohr frequencies $\omega$ (see Appendix~\ref{app:weak} for details).
We factor out the coupling strength $\lambda$ such that $A_\alpha= \mc{O}(1)$ (independent of $\lambda$).

We denote the size of the free system Hamiltonian $H$ by $h$ (measuring the size of the smallest energy gap and not to be confused with the Planck constant) and of the perturbing Hamiltonian by $v$ (taking $V=\mc{O}(1)$).
The important timescales are those of the intrinsic system dynamics $\tau_S \sim h^{-1}$, the perturbation $\tau_V \sim v^{-1}$, the system relaxation $\tau_R \sim \lambda^{-2} \gamma^{-1}$, and the bath correlation decay $\tau_B$.
We make the following assumptions: i) Born-Markov approximation, $\tau_B \ll \tau_R$, ii) rotating wave approximation, $\tau_S \ll \tau_R$,~\cite{Breuer2007Theory,lidar2019lecture} iii) small perturbation relative to the bath, $\tau_B \ll \tau_V$, and iv) small perturbation relative to the system, $\tau_S \ll \tau_V$.

Upon perturbing $H \to H' = H + vV$, we replace $H_\mathrm{LS} \to H'_\mathrm{LS} = H_\mathrm{LS} + v H^{(1)}_\mathrm{LS}$ and $\mc{D} \to \mc{D}' = \mc{D} + v \mc{D}^{(1)}$ to first order in $v$.
This alters the Bohr frequencies $\omega$ and components $A_\alpha(\omega)$.
Expressions for these are derived in Appendix~\ref{app:weak}, the perturbations $H^{(1)}_\mathrm{LS},\, \mc{D}^{(1)}$ being of size $\epsilon := \max_\psi \|\mc{H}_\mathrm{LS}^{(1)}(\psi) + \mc{D}^{(1)}(\psi)\| = \mc{O}(\tau_S / \tau_R) + \mc{O}(\tau_B / \tau_R)$.
It follows from assumptions (i)-(iv) that these terms are small compared with others in the master equation.

Applying bound~\eqref{eqn:main_bound} to this setting, we identify the \emph{true} perturbed trajectory $\frac{\dd \eta_t}{\dd t} = \mc{L}'(\eta_t)$ and the \emph{approximate} perturbed trajectory $\frac{\dd \sigma_t}{\dd t} = \mc{L}(\sigma_t) + v \mc{V}(\sigma_t)$.
In the latter, we only perturb the Hamiltonian term and ignore additional terms of size $\epsilon$.
All trajectories have the same initial state $\rho_0$.

\begin{res} \label{res:weak_coupling}
    For an open system in the weak coupling regime perturbed by the Hamiltonian $vV$,
    \begin{equation} \label{eqn:weak_coupling_speed}
        \bures(\rho_t, \eta_t) \leq \frac{1}{2} \int_0^t \dd s \; v \sqrt{\mc{F}(\eta_s, \mc{V})} + \Delta(t),
    \end{equation}
    where the error term is bounded by the estimate
    \begin{equation} \label{eqn:weak_coupling_error}
        \abs{\Delta(t)} \lesssim \Delta_\mathrm{est}(t) := \frac{4\sqrt{2}}{3} \|V\| \epsilon^\frac{1}{2} (vt)^\frac{3}{2} + \epsilon v t.
    \end{equation}
\end{res}
See Appendix~\ref{app:errors} for the proof.
For short times, the QFI term in Eq.~\eqref{eqn:weak_coupling_speed} is roughly $v t \sqrt{\mc{F}(\rho_0,\mc{V})}$ -- hence, the error is negligible when $\sqrt{\mc{F}(\rho_0,\mc{V})} \gg \max\{\sqrt{\epsilon v t}, \, \epsilon\}$.

In specific cases, one can determine the error parameter $\epsilon$ more precisely.
We demonstrate this for a spin-boson model of two qubits interacting with a bath of many harmonic oscillators~\cite{lidar2019lecture}.
We take $H = \frac{h}{2}(\sigma_z \ox \id + \id \ox \sigma_z)$, $V = \frac{1}{2}(\sigma_x \ox \id + \id \ox \sigma_x)$, and an independent coupling of each qubit to a bath of the form $\lambda \sigma_z \ox \sum_k g_k (b_k + b_k^\dagger)$, $g_k$ being dimensionless coefficients.
Here, $\sigma_i$ are Pauli matrices and $b_k$ is the annihilation operator for the bosonic mode $k$.
This gives local dephasing dynamics $\mc{D}(\rho) = \lambda^2 \gamma \left[ (\sigma_z \ox \id) \rho (\sigma_z \ox \id) + (\id \ox \sigma_z ) \rho (\id  \ox \sigma_z) - 2 \rho \right]$, writing $\gamma = \gamma(0)$.
Then we find $\epsilon \leq 4 \lambda^2 \gamma / h$ (see Appendix~\ref{app:qubits}).
In this case, the component of order $\tau_B/\tau_R$ vanishes.
A numerical demonstration of the tightness of the speed limit for this example is shown in Fig.~\ref{fig:qubits}.\\

\inlineheading{Witnessing large QFI}
Quantum correlations and other resources may be witnessed experimentally by showing that the QFI under some Hamiltonian $H$ exceeds a given threshold $\mc{F}_*$.
For closed systems, a standard method~\cite{Frowis2016Lower} obtains a lower bound to $\bures(\rho_0,\rho_t)$ with evolution under $V$, by measuring the system at either time $0$ or $t$.
Each measurement has probability distribution $p_i(0),\, p_i(t)$; their similarity is quantified by the Bhattacharyya coefficient~\cite{Bhattacharyya1943Measure} $B(\vb{p}(0), \vb{p}(t)) := \sum_i \sqrt{p_i(0) p_i(t)}$, which satisfies $\arccos B(\vb{p}(0), \vb{p}(t)) \leq \bures(\rho_0,\rho_t)$.
This bound holds for any measurement and can be saturated.
Thanks to Eq.~\eqref{eqn:uhlmann_bound}, the resource is thus witnessed when $\frac{2}{tv} \arccos B(\vb{p}(0),\vb{p}(t)) > \sqrt{\mc{F}_*}$.

This standard method neglects decoherence, so we propose a protocol for open systems.
The idea is to measure the system, with or without perturbation, for a known time, and use the distinguishability of the trajectories to lower-bound the average speed of response via Results \ref{res:main} and \ref{res:weak_coupling}.
In two types of experimental runs, either the system evolves under the free dynamics, or one adds the perturbation $v \mc{V}$.
In each case, the same measurement is performed at a known time $t$, giving statistics $p_i(t)$ and $q_i(t)$ respectively.
First assuming the perturbation is exactly $\mc{P} = v \mc{V}$, the right-hand side of Eq.~\eqref{eqn:main_bound} is $\frac{tv}{2}$ times the time-averaged value of $\sqrt{\mc{F}(\sigma_s,\mc{V})}$, which quantifies the average speed of response to the perturbation.
It follows that the resource must be present at some time $s \in [0,t]$ along the perturbed trajectory whenever
\begin{equation} \label{eqn:witness}
    \frac{2\arccos B(\vb{p}(t),\vb{q}(t))}{t v} > \sqrt{\mc{F}_*},
\end{equation}
as the threshold $\mc{F}_*$ is exceeded.
In Appendix~\ref{app:main_derivation} we generalise this to a time-varying coefficient $v_t$.
In weak coupling, the error $\Delta_\mathrm{est}(t)$ from Eq.~\eqref{eqn:weak_coupling_speed} increases the threshold in Eq.~\eqref{eqn:witness} to $\sqrt{\mc{F}_*} + \frac{2\Delta_\mathrm{est}(t)}{vt}$.
The change in this threshold is $\mc{O}( \sqrt{\epsilon v t}) + \mc{O}(\epsilon)$.

A demonstration for witnessing entanglement is shown in Fig.~\ref{fig:qubits} for the two-qubit dephasing model described above, taking a Bell-basis measurement for the Bhattacharrya coefficient.
For any two-qubit separable state $\rho_\mathrm{sep}$ with the chosen local $V$, we have $\mc{F}(\rho_\mathrm{sep}, \mc{V}) \leq \mc{F}_* = 2$~\cite{Pezze2009Entanglement,Toth2012Multipartite,Hyllus2012Fisher}.
Here, this threshold is broken by the exact QFI for $t \lessapprox 1.41$, while entanglement is witnessed taking into account the error estimate for $t \lessapprox 1.26$.\\

\inlineheading{Quantum work fluctuations}
Here, we show implications for driving a system out of equilibrium.
Consider a system with Hamiltonian $H$ initially in thermal equilibrium at inverse temperature $\beta$, in the Gibbs state $\gibbs = \frac{e^{-\beta H}}{\tr e^{-\beta H}}$.
At time $0$, $H$ is quickly changed to $H'$, involving fluctuating work $w$ done on the system.
The mean and variance of work are computed from $\Delta H := H'-H$: $\ev{w} = \tr[\gibbs \Delta  H], \, \variance_w = \variance(\gibbs, \Delta H)$.
If the system is left to thermalise to the new Gibbs state $\gibbs' = \frac{e^{-\beta H'}}{\tr e^{-\beta H'}}$, then its Helmholtz free energy $F_{H,\beta}$ decreases.
This is defined by $F_{H,\beta} = \tr[\gibbs H] - \beta^{-1} S(\gibbs)$, where $S(\gibbs) = -\tr[\gibbs \ln \gibbs]$ is the von Neumann entropy~\cite{Goold2016Role}.
The second law of thermodynamics implies that the change $\Delta F := F_{H',\beta} - F_{H,\beta} \leq \ev{w}$ -- 
this is equivalent to saying that the dissipated work $W_\mathrm{diss} := \ev{w} - \Delta F \geq 0$~\cite{Goold2016Role}.
$W_\mathrm{diss}$ is thus associated with nonequilibrium entropy production.

In order to study small deviations from equilibrium, we follow the paradigm of Refs.~\cite{Scandi2019Thermodynamic,Miller2019Work}, where $\Delta H$ is assumed small.
One finds a fluctuation-dissipation relation~\cite{Miller2019Work}
\begin{equation} \label{eqn:fdr}
    \frac{\beta}{2} \variance_w = W_\mathrm{diss} + Q_w.
\end{equation}
Here, $Q_w \geq 0$ is a quantum correction to the usual classical relation~\cite{Hermans1991Simple,Jarzynski1997Nonequilibrium}, thus Eq.~\eqref{eqn:fdr} represents a modification of a classical statistical law near equilibrium that takes into account quantum effects.
It also limits coherent protocols that aim to simultaneously minimise work fluctuations and dissipation~\cite{Miller2019Work}.

Eq.~\eqref{eqn:fdr} holds for various slow driving settings; in our case with a single small quench, $Q_w$ is determined by a quantity closely related to QFI:
\begin{equation} \label{eqn:qw_def}
    Q_w = \frac{\beta}{2} \bar{I}(\gibbs, \Delta H),
\end{equation}
where $\bar{I}(\rho,A) := \int_0^1 \dd k \; \frac{1}{2} \tr\left( [\rho^k, A][A, \rho^{1-k}] \right)$.
The details of this result~\cite{Miller2019Work} are recalled in Appendix~\ref{app:work}.
$\bar{I}$ belongs to a family of generalised QFI quantities~\cite{Petz2002Covariance} whose members are interpreted as measures of quantum coherence (also known as asymmetry in this context): $\bar{I}(\rho, H)$ and $\mc{F}(\rho, \mc{H})$, among others, quantify the coherence of a state $\rho$ with respect to a Hamiltonian $H$~\cite{Yadin2016General,Girolami2014Observable,Zhang2017Detecting,Marvian2022Operational,Miller2018Energy}.
Moreover, they can be regarded as quantum contributions to the variance of $H$~\cite{Luo2005Quantum,Gibilisco2009Inequalities,Frerot2016Quantum}.
Some key properties justifying this interpretation are $\bar{I}(\rho,H) \leq \variance(\rho,H)$, with equality for pure states, and $\bar{I}(\rho,H)=0$ when $\rho$ commutes with $H$.
Therefore, as required of a measure of quantum work fluctuations, $2Q_w/\beta = \bar{I}(\gibbs,\Delta H)$ vanishes exactly when $[H,\Delta H]=0$.

\begin{res} \label{res:work}
    Quantum work fluctuations are necessary for fast departure from equilibrium.
    For a system weakly coupled to a thermal environment, at all times $t>0$ following the quench $H \to H'$, the distance between the initial state $\gibbs$ and the system's state $\rho_t$ obeys
    \begin{equation} \label{eqn:work_speed}
        \bures(\gibbs, \rho_t) \leq t \sqrt{3 \bar{I}(\gibbs, \Delta H)} + \Delta(t),
    \end{equation}
    where $\Delta(t)$ is the weak coupling error from Eq.~\eqref{eqn:weak_coupling_error}.
\end{res}
The proof is given in Appendix~\ref{app:work}.
Fast departure from $\gibbs$ thus requires large quantum work fluctuations as measured by $\bar{I}(\gibbs, \Delta H)$ -- equivalently, $\gibbs$ must have a high degree of quantum coherence with respect to $\Delta H$.

The physical importance of the correction $\Delta(t)$ is seen in the ``classical'' case where $[H, \Delta H]=0$ (i.e., the energy levels change but not the eigenstates).
Then $\bar{I}=0$, but the system must deviate from $\gibbs$ in order to reach the new steady state $\gibbs'$.
From our earlier discussion of the weak coupling error, by identifying $v$ with $\|\Delta H\|$, we therefore see that the \emph{quantum driving regime} -- when the $\bar{I}$ terms dominates on the right-hand side of Eq.~\eqref{eqn:work_speed} -- corresponds to $\frac{\sqrt{\bar{I}(\gibbs,\Delta H)}}{\|\Delta H\|} \gg  \max\{ \sqrt{\epsilon \|\Delta H\| t},\, \epsilon \}.$
The left-hand side of this inequality measures quantum work fluctuations relative to the size of $\Delta H$.
In the quantum driving regime, coherent evolution resulting from the changed Hamiltonian happens faster than thermalisation.\\

\inlineheading{Linear response}
Our results can also be applied to linear response theory, bounding the size of a response to a perturbation.
Following, for example, Ref.~\cite{Konopik2019Quantum}, assume the system begins at $t=0$ in a (possibly non-equilibrium) stationary state $\pi$, satisfying $\mc{L}(\pi)=0$.
The Hamiltonian is then perturbed by $v_t V$, giving the trajectory $\rho_t$.
The change in mean of $A$, $\delta A_t := \tr[A(\rho_t-\pi)] $, can be bounded from Eq.~\eqref{eqn:main_bound} by $\abs{\delta A_t} \leq \norm*{A} \int_0^t \dd s\, \sqrt{\mc{F}(\rho_s,\mc{P}_s)}$ (see Appendix~\ref{app:linear}).
In a weak coupling setting, approximating the QFI to lowest order in $v_s$, we have $\abs{\delta A_t} \lessapprox \norm*{A}\sqrt{\mc{F}(\pi,\mc{V})} \left[ \int_0^t \dd s\, \abs{v_s}  + \Delta(t) \right]$.
Note the term $\Delta(t)$ from Result~\ref{res:weak_coupling} reflecting the change in the dissipator~\cite{Levy2021Response}.
Alternatively, the observable speed limit Eq.~\eqref{eqn:obs_lim} similarly implies $\abs{\delta A_t} \lessapprox \sqrt{\variance(\pi,A) \mc{F}(\pi,\mc{V})} \left[\int_0^t \dd s\, \abs{v_s} + \Delta(t) \right]$. 
This requires the additional approximation that $D_{\tr}(\rho_t,\pi)$ is small for short times, but may give a tighter bound in replacing $\norm*{A}$ by the generally smaller $\sqrt{\variance(\pi,A)}$.
\\

\inlineheading{Outlook}
In summary, we have shown that the Mandelstam-Tamm speed limit can be extended to describe the response of an open system to a perturbation, with applications to quantum resource witnessing and thermodynamics.
There are several possible future directions.
Firstly, note that we can derive a similar speed limit replacing the Bures angle by the quantity $\tilde{\theta}(\rho,\sigma) = \arccos \tr(\sqrt{\rho}\sqrt{\sigma})$ and the QFI by (four times) the Wigner-Yanase skew information~\cite{Gibilisco2003Wigner} $I^\mathrm{WY}(\rho,\mc{V}) = -\frac{1}{2} \tr([\sqrt{\rho},V]^2)$ -- is this possible for generalised QFI quantities~\cite{Petz2002Covariance}?
Secondly, for weak coupling, it would be interesting to consider slowly varying perturbations with adiabatic master equations~\cite{Albash2012Quantum}, and implications for thermodynamic uncertainty relations which relate current fluctuations to entropy production~\cite{Horowitz2020Thermodynamic}.
Additionally, since our speed limit holds under Markovian dynamics, a violation might be used as a witness of non-Markovianity.
This would add to a library of existing witnesses, including those based on monotonic decrease of QFI~\cite{Rivas2014Quantum,Scandi2019Thermodynamic}.
A range of other interesting applications involves studying the rate at which a perturbation can generate resources, generalising past approaches by allowing for background decoherence.
This covers charging quantum batteries~\cite{Julia-Farre2020Bounds,Gyhm2022Quantum}, where one could include thermalisation during charging, and the production of quantum resources such as coherence and entanglement~\cite{Mohan2022Quantum}. 
\\

\begin{acknowledgments}
We thank Florian Fr\"owis, George Knee, and Bill Munro for discussions that inspired this work, as well as Stefan Nimmrichter for further input.
This project has received funding from the European Union's Horizon 2020 research and innovation programme under the Marie Sk\l{}odowska-Curie grant agreement No.~945422,
the DAAD,
the Deutsche Forschungsgemeinschaft (DFG, German Research Foundation, project numbers 447948357 and 440958198),
the Sino-German Center for Research Promotion (Project M-0294),
the ERC (Consolidator Grant 683107/TempoQ), and 
the German Ministry of Education and Research (Project QuKuK, BMBF Grant No. 16KIS1618K).
\end{acknowledgments}

\nocite{Hubner1992Explicit,Augusiak2016Asymptotic,Horn1985Matrix,Buca2012Note,sakurai1995modern,Kempe2001Theory,Kubo1957Statistical}

\bibliography{main}

\begin{thebibliography}{84}%
\makeatletter
\providecommand \@ifxundefined [1]{%
 \@ifx{#1\undefined}
}%
\providecommand \@ifnum [1]{%
 \ifnum #1\expandafter \@firstoftwo
 \else \expandafter \@secondoftwo
 \fi
}%
\providecommand \@ifx [1]{%
 \ifx #1\expandafter \@firstoftwo
 \else \expandafter \@secondoftwo
 \fi
}%
\providecommand \natexlab [1]{#1}%
\providecommand \enquote  [1]{``#1''}%
\providecommand \bibnamefont  [1]{#1}%
\providecommand \bibfnamefont [1]{#1}%
\providecommand \citenamefont [1]{#1}%
\providecommand \href@noop [0]{\@secondoftwo}%
\providecommand \href [0]{\begingroup \@sanitize@url \@href}%
\providecommand \@href[1]{\@@startlink{#1}\@@href}%
\providecommand \@@href[1]{\endgroup#1\@@endlink}%
\providecommand \@sanitize@url [0]{\catcode `\\12\catcode `\$12\catcode
  `\&12\catcode `\#12\catcode `\^12\catcode `\_12\catcode `\%12\relax}%
\providecommand \@@startlink[1]{}%
\providecommand \@@endlink[0]{}%
\providecommand \url  [0]{\begingroup\@sanitize@url \@url }%
\providecommand \@url [1]{\endgroup\@href {#1}{\urlprefix }}%
\providecommand \urlprefix  [0]{URL }%
\providecommand \Eprint [0]{\href }%
\providecommand \doibase [0]{https://doi.org/}%
\providecommand \selectlanguage [0]{\@gobble}%
\providecommand \bibinfo  [0]{\@secondoftwo}%
\providecommand \bibfield  [0]{\@secondoftwo}%
\providecommand \translation [1]{[#1]}%
\providecommand \BibitemOpen [0]{}%
\providecommand \bibitemStop [0]{}%
\providecommand \bibitemNoStop [0]{.\EOS\space}%
\providecommand \EOS [0]{\spacefactor3000\relax}%
\providecommand \BibitemShut  [1]{\csname bibitem#1\endcsname}%
\let\auto@bib@innerbib\@empty
%</preamble>
\bibitem [{\citenamefont {Mandelstam}\ and\ \citenamefont
  {Tamm}(1945)}]{Mandelstam1945Uncertainty}%
  \BibitemOpen
  \bibfield  {author} {\bibinfo {author} {\bibfnamefont {L.}~\bibnamefont
  {Mandelstam}}\ and\ \bibinfo {author} {\bibfnamefont {I.}~\bibnamefont
  {Tamm}},\ }\bibfield  {title} {\bibinfo {title} {{The uncertainty relation
  between energy and time in non-relativistic quantum mechanics}},\ }\href@noop
  {} {\bibfield  {journal} {\bibinfo  {journal} {Journal of Physics (USSR)}\
  }\textbf {\bibinfo {volume} {IX}},\ \bibinfo {pages} {249} (\bibinfo {year}
  {1945})}\BibitemShut {NoStop}%
\bibitem [{\citenamefont {Margolus}\ and\ \citenamefont
  {Levitin}(1998)}]{Margolus1998Maximum}%
  \BibitemOpen
  \bibfield  {author} {\bibinfo {author} {\bibfnamefont {N.}~\bibnamefont
  {Margolus}}\ and\ \bibinfo {author} {\bibfnamefont {L.~B.}\ \bibnamefont
  {Levitin}},\ }\bibfield  {title} {\bibinfo {title} {{The maximum speed of
  dynamical evolution}},\ }\href
  {https://doi.org/10.1016/S0167-2789(98)00054-2} {\bibfield  {journal}
  {\bibinfo  {journal} {Physica D: Nonlinear Phenomena}\ }\textbf {\bibinfo
  {volume} {120}},\ \bibinfo {pages} {188} (\bibinfo {year}
  {1998})}\BibitemShut {NoStop}%
\bibitem [{\citenamefont {Jones}\ and\ \citenamefont
  {Kok}(2010)}]{Jones2010Geometric}%
  \BibitemOpen
  \bibfield  {author} {\bibinfo {author} {\bibfnamefont {P.~J.}\ \bibnamefont
  {Jones}}\ and\ \bibinfo {author} {\bibfnamefont {P.}~\bibnamefont {Kok}},\
  }\bibfield  {title} {\bibinfo {title} {{Geometric derivation of the quantum
  speed limit}},\ }\href {https://doi.org/10.1103/PhysRevA.82.022107}
  {\bibfield  {journal} {\bibinfo  {journal} {Physical Review A}\ }\textbf
  {\bibinfo {volume} {82}},\ \bibinfo {pages} {022107} (\bibinfo {year}
  {2010})}\BibitemShut {NoStop}%
\bibitem [{\citenamefont {del Campo}\ \emph {et~al.}(2013)\citenamefont {del
  Campo}, \citenamefont {Egusquiza}, \citenamefont {Plenio},\ and\
  \citenamefont {Huelga}}]{delCampo2013Quantum}%
  \BibitemOpen
  \bibfield  {author} {\bibinfo {author} {\bibfnamefont {A.}~\bibnamefont {del
  Campo}}, \bibinfo {author} {\bibfnamefont {I.~L.}\ \bibnamefont {Egusquiza}},
  \bibinfo {author} {\bibfnamefont {M.~B.}\ \bibnamefont {Plenio}},\ and\
  \bibinfo {author} {\bibfnamefont {S.~F.}\ \bibnamefont {Huelga}},\ }\bibfield
   {title} {\bibinfo {title} {{Quantum Speed Limits in Open System Dynamics}},\
  }\href {https://doi.org/10.1103/PhysRevLett.110.050403} {\bibfield  {journal}
  {\bibinfo  {journal} {Physical Review Letters}\ }\textbf {\bibinfo {volume}
  {110}},\ \bibinfo {pages} {050403} (\bibinfo {year} {2013})}\BibitemShut
  {NoStop}%
\bibitem [{\citenamefont {Pires}\ \emph {et~al.}(2016)\citenamefont {Pires},
  \citenamefont {Cianciaruso}, \citenamefont {C{\'{e}}leri}, \citenamefont
  {Adesso},\ and\ \citenamefont {Soares-Pinto}}]{Pires2016Generalized}%
  \BibitemOpen
  \bibfield  {author} {\bibinfo {author} {\bibfnamefont {D.~P.}\ \bibnamefont
  {Pires}}, \bibinfo {author} {\bibfnamefont {M.}~\bibnamefont {Cianciaruso}},
  \bibinfo {author} {\bibfnamefont {L.~C.}\ \bibnamefont {C{\'{e}}leri}},
  \bibinfo {author} {\bibfnamefont {G.}~\bibnamefont {Adesso}},\ and\ \bibinfo
  {author} {\bibfnamefont {D.~O.}\ \bibnamefont {Soares-Pinto}},\ }\bibfield
  {title} {\bibinfo {title} {{Generalized Geometric Quantum Speed Limits}},\
  }\href {https://doi.org/10.1103/PhysRevX.6.021031} {\bibfield  {journal}
  {\bibinfo  {journal} {Physical Review X}\ }\textbf {\bibinfo {volume} {6}},\
  \bibinfo {pages} {021031} (\bibinfo {year} {2016})}\BibitemShut {NoStop}%
\bibitem [{\citenamefont {Deffner}\ and\ \citenamefont
  {Lutz}(2013)}]{Deffner2013Quantum}%
  \BibitemOpen
  \bibfield  {author} {\bibinfo {author} {\bibfnamefont {S.}~\bibnamefont
  {Deffner}}\ and\ \bibinfo {author} {\bibfnamefont {E.}~\bibnamefont {Lutz}},\
  }\bibfield  {title} {\bibinfo {title} {{Quantum Speed Limit for Non-Markovian
  Dynamics}},\ }\href {https://doi.org/10.1103/PhysRevLett.111.010402}
  {\bibfield  {journal} {\bibinfo  {journal} {Physical Review Letters}\
  }\textbf {\bibinfo {volume} {111}},\ \bibinfo {pages} {010402} (\bibinfo
  {year} {2013})}\BibitemShut {NoStop}%
\bibitem [{\citenamefont {Marvian}\ \emph {et~al.}(2016)\citenamefont
  {Marvian}, \citenamefont {Spekkens},\ and\ \citenamefont
  {Zanardi}}]{Marvian2016Quantum}%
  \BibitemOpen
  \bibfield  {author} {\bibinfo {author} {\bibfnamefont {I.}~\bibnamefont
  {Marvian}}, \bibinfo {author} {\bibfnamefont {R.~W.}\ \bibnamefont
  {Spekkens}},\ and\ \bibinfo {author} {\bibfnamefont {P.}~\bibnamefont
  {Zanardi}},\ }\bibfield  {title} {\bibinfo {title} {{Quantum speed limits,
  coherence, and asymmetry}},\ }\href
  {https://doi.org/10.1103/PhysRevA.93.052331} {\bibfield  {journal} {\bibinfo
  {journal} {Physical Review A}\ }\textbf {\bibinfo {volume} {93}},\ \bibinfo
  {pages} {052331} (\bibinfo {year} {2016})}\BibitemShut {NoStop}%
\bibitem [{\citenamefont {Uhlmann}(1992)}]{Uhlmann1992Energy}%
  \BibitemOpen
  \bibfield  {author} {\bibinfo {author} {\bibfnamefont {A.}~\bibnamefont
  {Uhlmann}},\ }\bibfield  {title} {\bibinfo {title} {{An energy dispersion
  estimate}},\ }\href {https://doi.org/10.1016/0375-9601(92)90555-Z} {\bibfield
   {journal} {\bibinfo  {journal} {Physics Letters A}\ }\textbf {\bibinfo
  {volume} {161}},\ \bibinfo {pages} {329} (\bibinfo {year}
  {1992})}\BibitemShut {NoStop}%
\bibitem [{\citenamefont {Braunstein}\ and\ \citenamefont
  {Caves}(1994)}]{Braunstein1994Statistical}%
  \BibitemOpen
  \bibfield  {author} {\bibinfo {author} {\bibfnamefont {S.~L.}\ \bibnamefont
  {Braunstein}}\ and\ \bibinfo {author} {\bibfnamefont {C.~M.}\ \bibnamefont
  {Caves}},\ }\bibfield  {title} {\bibinfo {title} {{Statistical distance and
  the geometry of quantum states}},\ }\href
  {https://doi.org/10.1103/PhysRevLett.72.3439} {\bibfield  {journal} {\bibinfo
   {journal} {Physical Review Letters}\ }\textbf {\bibinfo {volume} {72}},\
  \bibinfo {pages} {3439} (\bibinfo {year} {1994})}\BibitemShut {NoStop}%
\bibitem [{\citenamefont {Fr{\"{o}}wis}(2012)}]{Frowis2012Kind}%
  \BibitemOpen
  \bibfield  {author} {\bibinfo {author} {\bibfnamefont {F.}~\bibnamefont
  {Fr{\"{o}}wis}},\ }\bibfield  {title} {\bibinfo {title} {{Kind of
  entanglement that speeds up quantum evolution}},\ }\href
  {https://doi.org/10.1103/PhysRevA.85.052127} {\bibfield  {journal} {\bibinfo
  {journal} {Physical Review A}\ }\textbf {\bibinfo {volume} {85}},\ \bibinfo
  {pages} {052127} (\bibinfo {year} {2012})}\BibitemShut {NoStop}%
\bibitem [{\citenamefont {Bremermann}(1965)}]{Bremermann1965Quantum}%
  \BibitemOpen
  \bibfield  {author} {\bibinfo {author} {\bibfnamefont {H.~J.}\ \bibnamefont
  {Bremermann}},\ }\bibfield  {title} {\bibinfo {title} {{Quantum noise and
  information}},\ }\href
  {http://books.google.com/books?hl=en&lr=&id=QsBQCBgTx8gC&oi=fnd&pg=PA15&dq=Quantum+noise+and+information&ots=qL5rzklhcy&sig=05HjZnznlJI1FFZRRoxHALrWVkM}
  {\bibfield  {journal} {\bibinfo  {journal} {Proceedings of the Fifth Berkeley
  Symposium on Mathematical Statistics and Probability}\ }\textbf {\bibinfo
  {volume} {222}},\ \bibinfo {pages} {15} (\bibinfo {year} {1965})}\BibitemShut
  {NoStop}%
\bibitem [{\citenamefont {Lloyd}(2000)}]{Lloyd2000Ultimate}%
  \BibitemOpen
  \bibfield  {author} {\bibinfo {author} {\bibfnamefont {S.}~\bibnamefont
  {Lloyd}},\ }\bibfield  {title} {\bibinfo {title} {{Ultimate physical limits
  to computation}},\ }\href {https://doi.org/10.1038/35023282} {\bibfield
  {journal} {\bibinfo  {journal} {Nature}\ }\textbf {\bibinfo {volume} {406}},\
  \bibinfo {pages} {1047} (\bibinfo {year} {2000})}\BibitemShut {NoStop}%
\bibitem [{\citenamefont {Deffner}(2020)}]{Deffner2020Quantum}%
  \BibitemOpen
  \bibfield  {author} {\bibinfo {author} {\bibfnamefont {S.}~\bibnamefont
  {Deffner}},\ }\bibfield  {title} {\bibinfo {title} {{Quantum speed limits and
  the maximal rate of information production}},\ }\href
  {https://doi.org/10.1103/PhysRevResearch.2.013161} {\bibfield  {journal}
  {\bibinfo  {journal} {Physical Review Research}\ }\textbf {\bibinfo {volume}
  {2}},\ \bibinfo {pages} {013161} (\bibinfo {year} {2020})}\BibitemShut
  {NoStop}%
\bibitem [{\citenamefont {Bekenstein}(1974)}]{Bekenstein1974Generalized}%
  \BibitemOpen
  \bibfield  {author} {\bibinfo {author} {\bibfnamefont {J.~D.}\ \bibnamefont
  {Bekenstein}},\ }\bibfield  {title} {\bibinfo {title} {{Generalized second
  law of thermodynamics in black-hole physics}},\ }\href
  {https://doi.org/10.1103/PhysRevD.9.3292} {\bibfield  {journal} {\bibinfo
  {journal} {Physical Review D}\ }\textbf {\bibinfo {volume} {9}},\ \bibinfo
  {pages} {3292} (\bibinfo {year} {1974})}\BibitemShut {NoStop}%
\bibitem [{\citenamefont {Deffner}\ and\ \citenamefont
  {Lutz}(2010)}]{Deffner2010Generalized}%
  \BibitemOpen
  \bibfield  {author} {\bibinfo {author} {\bibfnamefont {S.}~\bibnamefont
  {Deffner}}\ and\ \bibinfo {author} {\bibfnamefont {E.}~\bibnamefont {Lutz}},\
  }\bibfield  {title} {\bibinfo {title} {{Generalized Clausius Inequality for
  Nonequilibrium Quantum Processes}},\ }\href
  {https://doi.org/10.1103/PhysRevLett.105.170402} {\bibfield  {journal}
  {\bibinfo  {journal} {Physical Review Letters}\ }\textbf {\bibinfo {volume}
  {105}},\ \bibinfo {pages} {170402} (\bibinfo {year} {2010})}\BibitemShut
  {NoStop}%
\bibitem [{\citenamefont {Abah}\ and\ \citenamefont
  {Lutz}(2017)}]{Abah2017Energy}%
  \BibitemOpen
  \bibfield  {author} {\bibinfo {author} {\bibfnamefont {O.}~\bibnamefont
  {Abah}}\ and\ \bibinfo {author} {\bibfnamefont {E.}~\bibnamefont {Lutz}},\
  }\bibfield  {title} {\bibinfo {title} {{Energy efficient quantum machines}},\
  }\href {https://doi.org/10.1209/0295-5075/118/40005} {\bibfield  {journal}
  {\bibinfo  {journal} {EPL (Europhysics Letters)}\ }\textbf {\bibinfo {volume}
  {118}},\ \bibinfo {pages} {40005} (\bibinfo {year} {2017})}\BibitemShut
  {NoStop}%
\bibitem [{\citenamefont {Mukherjee}\ \emph {et~al.}(2016)\citenamefont
  {Mukherjee}, \citenamefont {Niedenzu}, \citenamefont {Kofman},\ and\
  \citenamefont {Kurizki}}]{Mukherjee2016Speed}%
  \BibitemOpen
  \bibfield  {author} {\bibinfo {author} {\bibfnamefont {V.}~\bibnamefont
  {Mukherjee}}, \bibinfo {author} {\bibfnamefont {W.}~\bibnamefont {Niedenzu}},
  \bibinfo {author} {\bibfnamefont {A.~G.}\ \bibnamefont {Kofman}},\ and\
  \bibinfo {author} {\bibfnamefont {G.}~\bibnamefont {Kurizki}},\ }\bibfield
  {title} {\bibinfo {title} {{Speed and efficiency limits of multilevel
  incoherent heat engines}},\ }\href
  {https://doi.org/10.1103/PhysRevE.94.062109} {\bibfield  {journal} {\bibinfo
  {journal} {Physical Review E}\ }\textbf {\bibinfo {volume} {94}},\ \bibinfo
  {pages} {062109} (\bibinfo {year} {2016})},\ \Eprint
  {https://arxiv.org/abs/1607.08452} {1607.08452} \BibitemShut {NoStop}%
\bibitem [{\citenamefont {Binder}\ \emph {et~al.}(2015)\citenamefont {Binder},
  \citenamefont {Vinjanampathy}, \citenamefont {Modi},\ and\ \citenamefont
  {Goold}}]{Binder2015Quantacell}%
  \BibitemOpen
  \bibfield  {author} {\bibinfo {author} {\bibfnamefont {F.~C.}\ \bibnamefont
  {Binder}}, \bibinfo {author} {\bibfnamefont {S.}~\bibnamefont
  {Vinjanampathy}}, \bibinfo {author} {\bibfnamefont {K.}~\bibnamefont
  {Modi}},\ and\ \bibinfo {author} {\bibfnamefont {J.}~\bibnamefont {Goold}},\
  }\bibfield  {title} {\bibinfo {title} {{Quantacell: powerful charging of
  quantum batteries}},\ }\href {https://doi.org/10.1088/1367-2630/17/7/075015}
  {\bibfield  {journal} {\bibinfo  {journal} {New Journal of Physics}\ }\textbf
  {\bibinfo {volume} {17}},\ \bibinfo {pages} {075015} (\bibinfo {year}
  {2015})}\BibitemShut {NoStop}%
\bibitem [{\citenamefont {Campaioli}\ \emph {et~al.}(2017)\citenamefont
  {Campaioli}, \citenamefont {Pollock}, \citenamefont {Binder}, \citenamefont
  {C{\'{e}}leri}, \citenamefont {Goold}, \citenamefont {Vinjanampathy},\ and\
  \citenamefont {Modi}}]{Campaioli2017Enhancing}%
  \BibitemOpen
  \bibfield  {author} {\bibinfo {author} {\bibfnamefont {F.}~\bibnamefont
  {Campaioli}}, \bibinfo {author} {\bibfnamefont {F.~A.}\ \bibnamefont
  {Pollock}}, \bibinfo {author} {\bibfnamefont {F.~C.}\ \bibnamefont {Binder}},
  \bibinfo {author} {\bibfnamefont {L.}~\bibnamefont {C{\'{e}}leri}}, \bibinfo
  {author} {\bibfnamefont {J.}~\bibnamefont {Goold}}, \bibinfo {author}
  {\bibfnamefont {S.}~\bibnamefont {Vinjanampathy}},\ and\ \bibinfo {author}
  {\bibfnamefont {K.}~\bibnamefont {Modi}},\ }\bibfield  {title} {\bibinfo
  {title} {{Enhancing the Charging Power of Quantum Batteries}},\ }\href
  {https://doi.org/10.1103/PhysRevLett.118.150601} {\bibfield  {journal}
  {\bibinfo  {journal} {Physical Review Letters}\ }\textbf {\bibinfo {volume}
  {118}},\ \bibinfo {pages} {150601} (\bibinfo {year} {2017})}\BibitemShut
  {NoStop}%
\bibitem [{\citenamefont {Mohan}\ and\ \citenamefont
  {Pati}(2022)}]{Mohan2022Quantum2}%
  \BibitemOpen
  \bibfield  {author} {\bibinfo {author} {\bibfnamefont {B.}~\bibnamefont
  {Mohan}}\ and\ \bibinfo {author} {\bibfnamefont {A.~K.}\ \bibnamefont
  {Pati}},\ }\bibfield  {title} {\bibinfo {title} {{Quantum speed limits for
  observables}},\ }\href {https://doi.org/10.1103/PhysRevA.106.042436}
  {\bibfield  {journal} {\bibinfo  {journal} {Physical Review A}\ }\textbf
  {\bibinfo {volume} {106}},\ \bibinfo {pages} {042436} (\bibinfo {year}
  {2022})}\BibitemShut {NoStop}%
\bibitem [{\citenamefont {Juli{\`{a}}-Farr{\'{e}}}\ \emph
  {et~al.}(2020)\citenamefont {Juli{\`{a}}-Farr{\'{e}}}, \citenamefont
  {Salamon}, \citenamefont {Riera}, \citenamefont {Bera},\ and\ \citenamefont
  {Lewenstein}}]{Julia-Farre2020Bounds}%
  \BibitemOpen
  \bibfield  {author} {\bibinfo {author} {\bibfnamefont {S.}~\bibnamefont
  {Juli{\`{a}}-Farr{\'{e}}}}, \bibinfo {author} {\bibfnamefont
  {T.}~\bibnamefont {Salamon}}, \bibinfo {author} {\bibfnamefont
  {A.}~\bibnamefont {Riera}}, \bibinfo {author} {\bibfnamefont {M.~N.}\
  \bibnamefont {Bera}},\ and\ \bibinfo {author} {\bibfnamefont
  {M.}~\bibnamefont {Lewenstein}},\ }\bibfield  {title} {\bibinfo {title}
  {{Bounds on the capacity and power of quantum batteries}},\ }\href
  {https://doi.org/10.1103/PhysRevResearch.2.023113} {\bibfield  {journal}
  {\bibinfo  {journal} {Physical Review Research}\ }\textbf {\bibinfo {volume}
  {2}},\ \bibinfo {pages} {023113} (\bibinfo {year} {2020})}\BibitemShut
  {NoStop}%
\bibitem [{\citenamefont {Gessner}\ and\ \citenamefont
  {Smerzi}(2018)}]{Gessner2018Statistical}%
  \BibitemOpen
  \bibfield  {author} {\bibinfo {author} {\bibfnamefont {M.}~\bibnamefont
  {Gessner}}\ and\ \bibinfo {author} {\bibfnamefont {A.}~\bibnamefont
  {Smerzi}},\ }\bibfield  {title} {\bibinfo {title} {{Statistical speed of
  quantum states: Generalized quantum Fisher information and Schatten speed}},\
  }\href {https://doi.org/10.1103/PhysRevA.97.022109} {\bibfield  {journal}
  {\bibinfo  {journal} {Physical Review A}\ }\textbf {\bibinfo {volume} {97}},\
  \bibinfo {pages} {022109} (\bibinfo {year} {2018})}\BibitemShut {NoStop}%
\bibitem [{\citenamefont {Zhang}\ \emph {et~al.}(2017)\citenamefont {Zhang},
  \citenamefont {Yadin}, \citenamefont {Hou}, \citenamefont {Cao},
  \citenamefont {Liu}, \citenamefont {Huang}, \citenamefont {Maity},
  \citenamefont {Vedral}, \citenamefont {Li}, \citenamefont {Guo},\ and\
  \citenamefont {Girolami}}]{Zhang2017Detecting}%
  \BibitemOpen
  \bibfield  {author} {\bibinfo {author} {\bibfnamefont {C.}~\bibnamefont
  {Zhang}}, \bibinfo {author} {\bibfnamefont {B.}~\bibnamefont {Yadin}},
  \bibinfo {author} {\bibfnamefont {Z.-B.}\ \bibnamefont {Hou}}, \bibinfo
  {author} {\bibfnamefont {H.}~\bibnamefont {Cao}}, \bibinfo {author}
  {\bibfnamefont {B.-H.}\ \bibnamefont {Liu}}, \bibinfo {author} {\bibfnamefont
  {Y.-F.}\ \bibnamefont {Huang}}, \bibinfo {author} {\bibfnamefont
  {R.}~\bibnamefont {Maity}}, \bibinfo {author} {\bibfnamefont
  {V.}~\bibnamefont {Vedral}}, \bibinfo {author} {\bibfnamefont {C.-F.}\
  \bibnamefont {Li}}, \bibinfo {author} {\bibfnamefont {G.-C.}\ \bibnamefont
  {Guo}},\ and\ \bibinfo {author} {\bibfnamefont {D.}~\bibnamefont
  {Girolami}},\ }\bibfield  {title} {\bibinfo {title} {{Detecting
  metrologically useful asymmetry and entanglement by a few local
  measurements}},\ }\href {https://doi.org/10.1103/PhysRevA.96.042327}
  {\bibfield  {journal} {\bibinfo  {journal} {Physical Review A}\ }\textbf
  {\bibinfo {volume} {96}},\ \bibinfo {pages} {042327} (\bibinfo {year}
  {2017})}\BibitemShut {NoStop}%
\bibitem [{\citenamefont {Paris}(2009)}]{Paris2009Quantum}%
  \BibitemOpen
  \bibfield  {author} {\bibinfo {author} {\bibfnamefont {M.~G.~A.}\
  \bibnamefont {Paris}},\ }\bibfield  {title} {\bibinfo {title} {{Quantum
  Estimation For Quantum Technology}},\ }\href
  {https://doi.org/10.1142/S0219749909004839} {\bibfield  {journal} {\bibinfo
  {journal} {International Journal of Quantum Information}\ }\textbf {\bibinfo
  {volume} {07}},\ \bibinfo {pages} {125} (\bibinfo {year} {2009})}\BibitemShut
  {NoStop}%
\bibitem [{\citenamefont {Amico}\ \emph {et~al.}(2008)\citenamefont {Amico},
  \citenamefont {Fazio}, \citenamefont {Osterloh},\ and\ \citenamefont
  {Vedral}}]{Amico2008Entanglement}%
  \BibitemOpen
  \bibfield  {author} {\bibinfo {author} {\bibfnamefont {L.}~\bibnamefont
  {Amico}}, \bibinfo {author} {\bibfnamefont {R.}~\bibnamefont {Fazio}},
  \bibinfo {author} {\bibfnamefont {A.}~\bibnamefont {Osterloh}},\ and\
  \bibinfo {author} {\bibfnamefont {V.}~\bibnamefont {Vedral}},\ }\bibfield
  {title} {\bibinfo {title} {{Entanglement in many-body systems}},\ }\href
  {https://doi.org/10.1103/RevModPhys.80.517} {\bibfield  {journal} {\bibinfo
  {journal} {Reviews of Modern Physics}\ }\textbf {\bibinfo {volume} {80}},\
  \bibinfo {pages} {517} (\bibinfo {year} {2008})}\BibitemShut {NoStop}%
\bibitem [{\citenamefont {Hyllus}\ \emph {et~al.}(2012)\citenamefont {Hyllus},
  \citenamefont {Laskowski}, \citenamefont {Krischek}, \citenamefont
  {Schwemmer}, \citenamefont {Wieczorek}, \citenamefont {Weinfurter},
  \citenamefont {Pezz{\'{e}}},\ and\ \citenamefont
  {Smerzi}}]{Hyllus2012Fisher}%
  \BibitemOpen
  \bibfield  {author} {\bibinfo {author} {\bibfnamefont {P.}~\bibnamefont
  {Hyllus}}, \bibinfo {author} {\bibfnamefont {W.}~\bibnamefont {Laskowski}},
  \bibinfo {author} {\bibfnamefont {R.}~\bibnamefont {Krischek}}, \bibinfo
  {author} {\bibfnamefont {C.}~\bibnamefont {Schwemmer}}, \bibinfo {author}
  {\bibfnamefont {W.}~\bibnamefont {Wieczorek}}, \bibinfo {author}
  {\bibfnamefont {H.}~\bibnamefont {Weinfurter}}, \bibinfo {author}
  {\bibfnamefont {L.}~\bibnamefont {Pezz{\'{e}}}},\ and\ \bibinfo {author}
  {\bibfnamefont {A.}~\bibnamefont {Smerzi}},\ }\bibfield  {title} {\bibinfo
  {title} {{Fisher information and multiparticle entanglement}},\ }\href
  {https://doi.org/10.1103/PhysRevA.85.022321} {\bibfield  {journal} {\bibinfo
  {journal} {Physical Review A}\ }\textbf {\bibinfo {volume} {85}},\ \bibinfo
  {pages} {022321} (\bibinfo {year} {2012})}\BibitemShut {NoStop}%
\bibitem [{\citenamefont {T{\'{o}}th}(2012)}]{Toth2012Multipartite}%
  \BibitemOpen
  \bibfield  {author} {\bibinfo {author} {\bibfnamefont {G.}~\bibnamefont
  {T{\'{o}}th}},\ }\bibfield  {title} {\bibinfo {title} {{Multipartite
  entanglement and high-precision metrology}},\ }\href
  {https://doi.org/10.1103/PhysRevA.85.022322} {\bibfield  {journal} {\bibinfo
  {journal} {Physical Review A}\ }\textbf {\bibinfo {volume} {85}},\ \bibinfo
  {pages} {022322} (\bibinfo {year} {2012})}\BibitemShut {NoStop}%
\bibitem [{\citenamefont {Morris}\ \emph {et~al.}(2020)\citenamefont {Morris},
  \citenamefont {Yadin}, \citenamefont {Fadel}, \citenamefont {Zibold},
  \citenamefont {Treutlein},\ and\ \citenamefont
  {Adesso}}]{Morris2020Entanglement}%
  \BibitemOpen
  \bibfield  {author} {\bibinfo {author} {\bibfnamefont {B.}~\bibnamefont
  {Morris}}, \bibinfo {author} {\bibfnamefont {B.}~\bibnamefont {Yadin}},
  \bibinfo {author} {\bibfnamefont {M.}~\bibnamefont {Fadel}}, \bibinfo
  {author} {\bibfnamefont {T.}~\bibnamefont {Zibold}}, \bibinfo {author}
  {\bibfnamefont {P.}~\bibnamefont {Treutlein}},\ and\ \bibinfo {author}
  {\bibfnamefont {G.}~\bibnamefont {Adesso}},\ }\bibfield  {title} {\bibinfo
  {title} {{Entanglement between Identical Particles Is a Useful and Consistent
  Resource}},\ }\href {https://doi.org/10.1103/PhysRevX.10.041012} {\bibfield
  {journal} {\bibinfo  {journal} {Physical Review X}\ }\textbf {\bibinfo
  {volume} {10}},\ \bibinfo {pages} {041012} (\bibinfo {year}
  {2020})}\BibitemShut {NoStop}%
\bibitem [{\citenamefont {Pezz{\'{e}}}\ and\ \citenamefont
  {Smerzi}(2009)}]{Pezze2009Entanglement}%
  \BibitemOpen
  \bibfield  {author} {\bibinfo {author} {\bibfnamefont {L.}~\bibnamefont
  {Pezz{\'{e}}}}\ and\ \bibinfo {author} {\bibfnamefont {A.}~\bibnamefont
  {Smerzi}},\ }\bibfield  {title} {\bibinfo {title} {{Entanglement, Nonlinear
  Dynamics, and the Heisenberg Limit}},\ }\href
  {https://doi.org/10.1103/PhysRevLett.102.100401} {\bibfield  {journal}
  {\bibinfo  {journal} {Physical Review Letters}\ }\textbf {\bibinfo {volume}
  {102}},\ \bibinfo {pages} {100401} (\bibinfo {year} {2009})}\BibitemShut
  {NoStop}%
\bibitem [{\citenamefont {Pezz{\`{e}}}\ \emph {et~al.}(2016)\citenamefont
  {Pezz{\`{e}}}, \citenamefont {Li}, \citenamefont {Li},\ and\ \citenamefont
  {Smerzi}}]{Pezze2015Witnessing}%
  \BibitemOpen
  \bibfield  {author} {\bibinfo {author} {\bibfnamefont {L.}~\bibnamefont
  {Pezz{\`{e}}}}, \bibinfo {author} {\bibfnamefont {Y.}~\bibnamefont {Li}},
  \bibinfo {author} {\bibfnamefont {W.}~\bibnamefont {Li}},\ and\ \bibinfo
  {author} {\bibfnamefont {A.}~\bibnamefont {Smerzi}},\ }\bibfield  {title}
  {\bibinfo {title} {{Witnessing entanglement without entanglement witness
  operators}},\ }\href {https://doi.org/10.1073/pnas.1603346113} {\bibfield
  {journal} {\bibinfo  {journal} {Proceedings of the National Academy of
  Sciences}\ }\textbf {\bibinfo {volume} {113}},\ \bibinfo {pages} {11459}
  (\bibinfo {year} {2016})}\BibitemShut {NoStop}%
\bibitem [{\citenamefont {Yadin}\ \emph {et~al.}(2021)\citenamefont {Yadin},
  \citenamefont {Fadel},\ and\ \citenamefont
  {Gessner}}]{Yadin2021Metrological}%
  \BibitemOpen
  \bibfield  {author} {\bibinfo {author} {\bibfnamefont {B.}~\bibnamefont
  {Yadin}}, \bibinfo {author} {\bibfnamefont {M.}~\bibnamefont {Fadel}},\ and\
  \bibinfo {author} {\bibfnamefont {M.}~\bibnamefont {Gessner}},\ }\bibfield
  {title} {\bibinfo {title} {{Metrological complementarity reveals the
  Einstein-Podolsky-Rosen paradox}},\ }\href
  {https://doi.org/10.1038/s41467-021-22353-3} {\bibfield  {journal} {\bibinfo
  {journal} {Nature Communications}\ }\textbf {\bibinfo {volume} {12}},\
  \bibinfo {pages} {2410} (\bibinfo {year} {2021})}\BibitemShut {NoStop}%
\bibitem [{\citenamefont {Yadin}\ and\ \citenamefont
  {Vedral}(2016)}]{Yadin2016General}%
  \BibitemOpen
  \bibfield  {author} {\bibinfo {author} {\bibfnamefont {B.}~\bibnamefont
  {Yadin}}\ and\ \bibinfo {author} {\bibfnamefont {V.}~\bibnamefont {Vedral}},\
  }\bibfield  {title} {\bibinfo {title} {{General framework for quantum
  macroscopicity in terms of coherence}},\ }\href
  {https://doi.org/10.1103/PhysRevA.93.022122} {\bibfield  {journal} {\bibinfo
  {journal} {Physical Review A}\ }\textbf {\bibinfo {volume} {93}},\ \bibinfo
  {pages} {022122} (\bibinfo {year} {2016})}\BibitemShut {NoStop}%
\bibitem [{\citenamefont {Marvian}(2020)}]{Marvian2020Coherence}%
  \BibitemOpen
  \bibfield  {author} {\bibinfo {author} {\bibfnamefont {I.}~\bibnamefont
  {Marvian}},\ }\bibfield  {title} {\bibinfo {title} {{Coherence distillation
  machines are impossible in quantum thermodynamics}},\ }\href
  {https://doi.org/10.1038/s41467-019-13846-3} {\bibfield  {journal} {\bibinfo
  {journal} {Nature Communications}\ }\textbf {\bibinfo {volume} {11}},\
  \bibinfo {pages} {25} (\bibinfo {year} {2020})},\ \Eprint
  {https://arxiv.org/abs/1805.01989} {1805.01989} \BibitemShut {NoStop}%
\bibitem [{\citenamefont {Marvian}(2022)}]{Marvian2022Operational}%
  \BibitemOpen
  \bibfield  {author} {\bibinfo {author} {\bibfnamefont {I.}~\bibnamefont
  {Marvian}},\ }\bibfield  {title} {\bibinfo {title} {{Operational
  Interpretation of Quantum Fisher Information in Quantum Thermodynamics}},\
  }\href {https://doi.org/10.1103/PhysRevLett.129.190502} {\bibfield  {journal}
  {\bibinfo  {journal} {Physical Review Letters}\ }\textbf {\bibinfo {volume}
  {129}},\ \bibinfo {pages} {190502} (\bibinfo {year} {2022})}\BibitemShut
  {NoStop}%
\bibitem [{\citenamefont {Fr{\"{o}}wis}\ \emph {et~al.}(2018)\citenamefont
  {Fr{\"{o}}wis}, \citenamefont {Sekatski}, \citenamefont {D{\"{u}}r},
  \citenamefont {Gisin},\ and\ \citenamefont
  {Sangouard}}]{Frowis2018Macroscopic}%
  \BibitemOpen
  \bibfield  {author} {\bibinfo {author} {\bibfnamefont {F.}~\bibnamefont
  {Fr{\"{o}}wis}}, \bibinfo {author} {\bibfnamefont {P.}~\bibnamefont
  {Sekatski}}, \bibinfo {author} {\bibfnamefont {W.}~\bibnamefont {D{\"{u}}r}},
  \bibinfo {author} {\bibfnamefont {N.}~\bibnamefont {Gisin}},\ and\ \bibinfo
  {author} {\bibfnamefont {N.}~\bibnamefont {Sangouard}},\ }\bibfield  {title}
  {\bibinfo {title} {{Macroscopic quantum states: Measures, fragility, and
  implementations}},\ }\href {https://doi.org/10.1103/RevModPhys.90.025004}
  {\bibfield  {journal} {\bibinfo  {journal} {Reviews of Modern Physics}\
  }\textbf {\bibinfo {volume} {90}},\ \bibinfo {pages} {025004} (\bibinfo
  {year} {2018})}\BibitemShut {NoStop}%
\bibitem [{\citenamefont {Yadin}\ \emph {et~al.}(2018)\citenamefont {Yadin},
  \citenamefont {Binder}, \citenamefont {Thompson}, \citenamefont
  {Narasimhachar}, \citenamefont {Gu},\ and\ \citenamefont
  {Kim}}]{Yadin2018Operational}%
  \BibitemOpen
  \bibfield  {author} {\bibinfo {author} {\bibfnamefont {B.}~\bibnamefont
  {Yadin}}, \bibinfo {author} {\bibfnamefont {F.~C.}\ \bibnamefont {Binder}},
  \bibinfo {author} {\bibfnamefont {J.}~\bibnamefont {Thompson}}, \bibinfo
  {author} {\bibfnamefont {V.}~\bibnamefont {Narasimhachar}}, \bibinfo {author}
  {\bibfnamefont {M.}~\bibnamefont {Gu}},\ and\ \bibinfo {author}
  {\bibfnamefont {M.~S.}\ \bibnamefont {Kim}},\ }\bibfield  {title} {\bibinfo
  {title} {{Operational Resource Theory of Continuous-Variable
  Nonclassicality}},\ }\href {https://doi.org/10.1103/PhysRevX.8.041038}
  {\bibfield  {journal} {\bibinfo  {journal} {Physical Review X}\ }\textbf
  {\bibinfo {volume} {8}},\ \bibinfo {pages} {041038} (\bibinfo {year}
  {2018})}\BibitemShut {NoStop}%
\bibitem [{\citenamefont {Kwon}\ \emph {et~al.}(2019)\citenamefont {Kwon},
  \citenamefont {Tan}, \citenamefont {Volkoff},\ and\ \citenamefont
  {Jeong}}]{Kwon2019Nonclassicality}%
  \BibitemOpen
  \bibfield  {author} {\bibinfo {author} {\bibfnamefont {H.}~\bibnamefont
  {Kwon}}, \bibinfo {author} {\bibfnamefont {K.~C.}\ \bibnamefont {Tan}},
  \bibinfo {author} {\bibfnamefont {T.}~\bibnamefont {Volkoff}},\ and\ \bibinfo
  {author} {\bibfnamefont {H.}~\bibnamefont {Jeong}},\ }\bibfield  {title}
  {\bibinfo {title} {{Nonclassicality as a Quantifiable Resource for Quantum
  Metrology}},\ }\href {https://doi.org/10.1103/PhysRevLett.122.040503}
  {\bibfield  {journal} {\bibinfo  {journal} {Physical Review Letters}\
  }\textbf {\bibinfo {volume} {122}},\ \bibinfo {pages} {040503} (\bibinfo
  {year} {2019})}\BibitemShut {NoStop}%
\bibitem [{\citenamefont {Tan}\ \emph {et~al.}(2021)\citenamefont {Tan},
  \citenamefont {Narasimhachar},\ and\ \citenamefont {Regula}}]{Tan2021Fisher}%
  \BibitemOpen
  \bibfield  {author} {\bibinfo {author} {\bibfnamefont {K.~C.}\ \bibnamefont
  {Tan}}, \bibinfo {author} {\bibfnamefont {V.}~\bibnamefont {Narasimhachar}},\
  and\ \bibinfo {author} {\bibfnamefont {B.}~\bibnamefont {Regula}},\
  }\bibfield  {title} {\bibinfo {title} {{Fisher Information Universally
  Identifies Quantum Resources}},\ }\href
  {https://doi.org/10.1103/PhysRevLett.127.200402} {\bibfield  {journal}
  {\bibinfo  {journal} {Physical Review Letters}\ }\textbf {\bibinfo {volume}
  {127}},\ \bibinfo {pages} {200402} (\bibinfo {year} {2021})}\BibitemShut
  {NoStop}%
\bibitem [{\citenamefont {Girolami}(2014)}]{Girolami2014Observable}%
  \BibitemOpen
  \bibfield  {author} {\bibinfo {author} {\bibfnamefont {D.}~\bibnamefont
  {Girolami}},\ }\bibfield  {title} {\bibinfo {title} {{Observable Measure of
  Quantum Coherence in Finite Dimensional Systems}},\ }\href
  {https://doi.org/10.1103/PhysRevLett.113.170401} {\bibfield  {journal}
  {\bibinfo  {journal} {Physical Review Letters}\ }\textbf {\bibinfo {volume}
  {113}},\ \bibinfo {pages} {170401} (\bibinfo {year} {2014})}\BibitemShut
  {NoStop}%
\bibitem [{\citenamefont {Rath}\ \emph {et~al.}(2021)\citenamefont {Rath},
  \citenamefont {Branciard}, \citenamefont {Minguzzi},\ and\ \citenamefont
  {Vermersch}}]{Rath2021Quantum}%
  \BibitemOpen
  \bibfield  {author} {\bibinfo {author} {\bibfnamefont {A.}~\bibnamefont
  {Rath}}, \bibinfo {author} {\bibfnamefont {C.}~\bibnamefont {Branciard}},
  \bibinfo {author} {\bibfnamefont {A.}~\bibnamefont {Minguzzi}},\ and\
  \bibinfo {author} {\bibfnamefont {B.}~\bibnamefont {Vermersch}},\ }\bibfield
  {title} {\bibinfo {title} {{Quantum Fisher Information from Randomized
  Measurements}},\ }\href {https://doi.org/10.1103/PhysRevLett.127.260501}
  {\bibfield  {journal} {\bibinfo  {journal} {Physical Review Letters}\
  }\textbf {\bibinfo {volume} {127}},\ \bibinfo {pages} {260501} (\bibinfo
  {year} {2021})}\BibitemShut {NoStop}%
\bibitem [{\citenamefont {Strobel}\ \emph {et~al.}(2014)\citenamefont
  {Strobel}, \citenamefont {Muessel}, \citenamefont {Linnemann}, \citenamefont
  {Zibold}, \citenamefont {Hume}, \citenamefont {Pezze}, \citenamefont
  {Smerzi},\ and\ \citenamefont {Oberthaler}}]{Strobel2014Fisher}%
  \BibitemOpen
  \bibfield  {author} {\bibinfo {author} {\bibfnamefont {H.}~\bibnamefont
  {Strobel}}, \bibinfo {author} {\bibfnamefont {W.}~\bibnamefont {Muessel}},
  \bibinfo {author} {\bibfnamefont {D.}~\bibnamefont {Linnemann}}, \bibinfo
  {author} {\bibfnamefont {T.}~\bibnamefont {Zibold}}, \bibinfo {author}
  {\bibfnamefont {D.~B.}\ \bibnamefont {Hume}}, \bibinfo {author}
  {\bibfnamefont {L.}~\bibnamefont {Pezze}}, \bibinfo {author} {\bibfnamefont
  {A.}~\bibnamefont {Smerzi}},\ and\ \bibinfo {author} {\bibfnamefont {M.~K.}\
  \bibnamefont {Oberthaler}},\ }\bibfield  {title} {\bibinfo {title} {{Fisher
  information and entanglement of non-Gaussian spin states}},\ }\href
  {https://doi.org/10.1126/science.1250147} {\bibfield  {journal} {\bibinfo
  {journal} {Science}\ }\textbf {\bibinfo {volume} {345}},\ \bibinfo {pages}
  {424} (\bibinfo {year} {2014})}\BibitemShut {NoStop}%
\bibitem [{\citenamefont {Fr{\"{o}}wis}(2017)}]{Frowis2016Lower}%
  \BibitemOpen
  \bibfield  {author} {\bibinfo {author} {\bibfnamefont {F.}~\bibnamefont
  {Fr{\"{o}}wis}},\ }\bibfield  {title} {\bibinfo {title} {{Lower bounds on the
  size of general Schr{\"{o}}dinger-cat states from experimental data}},\
  }\href {https://doi.org/10.1088/1751-8121/aa5a92} {\bibfield  {journal}
  {\bibinfo  {journal} {Journal of Physics A: Mathematical and Theoretical}\
  }\textbf {\bibinfo {volume} {50}},\ \bibinfo {pages} {114003} (\bibinfo
  {year} {2017})}\BibitemShut {NoStop}%
\bibitem [{\citenamefont {Becker}\ \emph {et~al.}(2021)\citenamefont {Becker},
  \citenamefont {Datta}, \citenamefont {Lami},\ and\ \citenamefont
  {Rouz{\'{e}}}}]{Becker2021Energy}%
  \BibitemOpen
  \bibfield  {author} {\bibinfo {author} {\bibfnamefont {S.}~\bibnamefont
  {Becker}}, \bibinfo {author} {\bibfnamefont {N.}~\bibnamefont {Datta}},
  \bibinfo {author} {\bibfnamefont {L.}~\bibnamefont {Lami}},\ and\ \bibinfo
  {author} {\bibfnamefont {C.}~\bibnamefont {Rouz{\'{e}}}},\ }\bibfield
  {title} {\bibinfo {title} {{Energy-Constrained Discrimination of Unitaries,
  Quantum Speed Limits, and a Gaussian Solovay-Kitaev Theorem}},\ }\href
  {https://doi.org/10.1103/PhysRevLett.126.190504} {\bibfield  {journal}
  {\bibinfo  {journal} {Physical Review Letters}\ }\textbf {\bibinfo {volume}
  {126}},\ \bibinfo {pages} {190504} (\bibinfo {year} {2021})}\BibitemShut
  {NoStop}%
\bibitem [{\citenamefont {Suzuki}\ and\ \citenamefont
  {Takahashi}(2020)}]{Suzuki2020Performance}%
  \BibitemOpen
  \bibfield  {author} {\bibinfo {author} {\bibfnamefont {K.}~\bibnamefont
  {Suzuki}}\ and\ \bibinfo {author} {\bibfnamefont {K.}~\bibnamefont
  {Takahashi}},\ }\bibfield  {title} {\bibinfo {title} {{Performance evaluation
  of adiabatic quantum computation via quantum speed limits and possible
  applications to many-body systems}},\ }\href
  {https://doi.org/10.1103/PhysRevResearch.2.032016} {\bibfield  {journal}
  {\bibinfo  {journal} {Physical Review Research}\ }\textbf {\bibinfo {volume}
  {2}},\ \bibinfo {pages} {032016(R)} (\bibinfo {year} {2020})}\BibitemShut
  {NoStop}%
\bibitem [{\citenamefont {Hatomura}(2022)}]{Hatomura2022Performance}%
  \BibitemOpen
  \bibfield  {author} {\bibinfo {author} {\bibfnamefont {T.}~\bibnamefont
  {Hatomura}},\ }\bibfield  {title} {\bibinfo {title} {{Performance evaluation
  of invariant-based inverse engineering by quantum speed limits}},\ }\href
  {https://doi.org/10.1103/PhysRevA.106.L040401} {\bibfield  {journal}
  {\bibinfo  {journal} {Physical Review A}\ }\textbf {\bibinfo {volume}
  {106}},\ \bibinfo {pages} {L040401} (\bibinfo {year} {2022})}\BibitemShut
  {NoStop}%
\bibitem [{\citenamefont {Pandey}\ \emph {et~al.}(2023)\citenamefont {Pandey},
  \citenamefont {Bhowmick}, \citenamefont {Mohan}, \citenamefont {Sohail},\
  and\ \citenamefont {Sen}}]{Pandey2023Fundamental}%
  \BibitemOpen
  \bibfield  {author} {\bibinfo {author} {\bibfnamefont {V.}~\bibnamefont
  {Pandey}}, \bibinfo {author} {\bibfnamefont {S.}~\bibnamefont {Bhowmick}},
  \bibinfo {author} {\bibfnamefont {B.}~\bibnamefont {Mohan}}, \bibinfo
  {author} {\bibnamefont {Sohail}},\ and\ \bibinfo {author} {\bibfnamefont
  {U.}~\bibnamefont {Sen}},\ }\bibfield  {title} {\bibinfo {title}
  {{Fundamental speed limits on entanglement dynamics of bipartite quantum
  systems}},\ }\href {http://arxiv.org/abs/2303.07415} {\bibfield  {journal}
  {\bibinfo  {journal} {arXiv:2303.07415}\ } (\bibinfo {year}
  {2023})}\BibitemShut {NoStop}%
\bibitem [{\citenamefont {Albarelli}\ and\ \citenamefont
  {Demkowicz-Dobrza{\'{n}}ski}(2022)}]{Albarelli2022Probe}%
  \BibitemOpen
  \bibfield  {author} {\bibinfo {author} {\bibfnamefont {F.}~\bibnamefont
  {Albarelli}}\ and\ \bibinfo {author} {\bibfnamefont {R.}~\bibnamefont
  {Demkowicz-Dobrza{\'{n}}ski}},\ }\bibfield  {title} {\bibinfo {title} {{Probe
  Incompatibility in Multiparameter Noisy Quantum Metrology}},\ }\href
  {https://doi.org/10.1103/PhysRevX.12.011039} {\bibfield  {journal} {\bibinfo
  {journal} {Physical Review X}\ }\textbf {\bibinfo {volume} {12}},\ \bibinfo
  {pages} {011039} (\bibinfo {year} {2022})}\BibitemShut {NoStop}%
\bibitem [{\citenamefont {Gorini}(1976)}]{Gorini1976Completely}%
  \BibitemOpen
  \bibfield  {author} {\bibinfo {author} {\bibfnamefont {V.}~\bibnamefont
  {Gorini}},\ }\bibfield  {title} {\bibinfo {title} {{Completely positive
  dynamical semigroups of N-level systems}},\ }\href
  {https://doi.org/10.1063/1.522979} {\bibfield  {journal} {\bibinfo  {journal}
  {Journal of Mathematical Physics}\ }\textbf {\bibinfo {volume} {17}},\
  \bibinfo {pages} {821} (\bibinfo {year} {1976})}\BibitemShut {NoStop}%
\bibitem [{\citenamefont {Lindblad}(1976)}]{Lindblad1979On}%
  \BibitemOpen
  \bibfield  {author} {\bibinfo {author} {\bibfnamefont {G.}~\bibnamefont
  {Lindblad}},\ }\bibfield  {title} {\bibinfo {title} {{On the generators of
  quantum dynamical semigroups}},\ }\href {https://doi.org/10.1007/BF01608499}
  {\bibfield  {journal} {\bibinfo  {journal} {Communications in Mathematical
  Physics}\ }\textbf {\bibinfo {volume} {48}},\ \bibinfo {pages} {119}
  (\bibinfo {year} {1976})}\BibitemShut {NoStop}%
\bibitem [{\citenamefont {Rivas}\ \emph {et~al.}(2014)\citenamefont {Rivas},
  \citenamefont {Huelga},\ and\ \citenamefont {Plenio}}]{Rivas2014Quantum}%
  \BibitemOpen
  \bibfield  {author} {\bibinfo {author} {\bibfnamefont {{\'{A}}.}~\bibnamefont
  {Rivas}}, \bibinfo {author} {\bibfnamefont {S.~F.}\ \bibnamefont {Huelga}},\
  and\ \bibinfo {author} {\bibfnamefont {M.~B.}\ \bibnamefont {Plenio}},\
  }\bibfield  {title} {\bibinfo {title} {{Quantum non-Markovianity:
  characterization, quantification and detection}},\ }\href
  {https://doi.org/10.1088/0034-4885/77/9/094001} {\bibfield  {journal}
  {\bibinfo  {journal} {Reports on Progress in Physics}\ }\textbf {\bibinfo
  {volume} {77}},\ \bibinfo {pages} {094001} (\bibinfo {year}
  {2014})}\BibitemShut {NoStop}%
\bibitem [{\citenamefont {Nielsen}\ and\ \citenamefont
  {Chuang}(2010)}]{NielsenChuang_ch9}%
  \BibitemOpen
  \bibfield  {author} {\bibinfo {author} {\bibfnamefont {M.~A.}\ \bibnamefont
  {Nielsen}}\ and\ \bibinfo {author} {\bibfnamefont {I.~L.}\ \bibnamefont
  {Chuang}},\ }\bibfield  {title} {\bibinfo {title} {{Quantum Computation and
  Quantum Information}}\ }(\bibinfo  {publisher} {Cambridge University Press},\
  \bibinfo {address} {Cambridge},\ \bibinfo {year} {2010})\ Chap.~\bibinfo
  {chapter} {9}\BibitemShut {NoStop}%
\bibitem [{\citenamefont {Taddei}\ \emph {et~al.}(2013)\citenamefont {Taddei},
  \citenamefont {Escher}, \citenamefont {Davidovich},\ and\ \citenamefont {{de
  Matos Filho}}}]{Taddei2013Quantum}%
  \BibitemOpen
  \bibfield  {author} {\bibinfo {author} {\bibfnamefont {M.~M.}\ \bibnamefont
  {Taddei}}, \bibinfo {author} {\bibfnamefont {B.~M.}\ \bibnamefont {Escher}},
  \bibinfo {author} {\bibfnamefont {L.}~\bibnamefont {Davidovich}},\ and\
  \bibinfo {author} {\bibfnamefont {R.~L.}\ \bibnamefont {{de Matos Filho}}},\
  }\bibfield  {title} {\bibinfo {title} {{Quantum Speed Limit for Physical
  Processes}},\ }\href {https://doi.org/10.1103/PhysRevLett.110.050402}
  {\bibfield  {journal} {\bibinfo  {journal} {Physical Review Letters}\
  }\textbf {\bibinfo {volume} {110}},\ \bibinfo {pages} {050402} (\bibinfo
  {year} {2013})}\BibitemShut {NoStop}%
\bibitem [{\citenamefont {Garc{\'{i}}a-Pintos}\ \emph
  {et~al.}(2022)\citenamefont {Garc{\'{i}}a-Pintos}, \citenamefont {Nicholson},
  \citenamefont {Green}, \citenamefont {del Campo},\ and\ \citenamefont
  {Gorshkov}}]{GarciaPintos2022Unifying}%
  \BibitemOpen
  \bibfield  {author} {\bibinfo {author} {\bibfnamefont {L.~P.}\ \bibnamefont
  {Garc{\'{i}}a-Pintos}}, \bibinfo {author} {\bibfnamefont {S.~B.}\
  \bibnamefont {Nicholson}}, \bibinfo {author} {\bibfnamefont {J.~R.}\
  \bibnamefont {Green}}, \bibinfo {author} {\bibfnamefont {A.}~\bibnamefont
  {del Campo}},\ and\ \bibinfo {author} {\bibfnamefont {A.~V.}\ \bibnamefont
  {Gorshkov}},\ }\bibfield  {title} {\bibinfo {title} {{Unifying Quantum and
  Classical Speed Limits on Observables}},\ }\href
  {https://doi.org/10.1103/PhysRevX.12.011038} {\bibfield  {journal} {\bibinfo
  {journal} {Physical Review X}\ }\textbf {\bibinfo {volume} {12}},\ \bibinfo
  {pages} {011038} (\bibinfo {year} {2022})}\BibitemShut {NoStop}%
\bibitem [{\citenamefont {Lidar}(2019)}]{lidar2019lecture}%
  \BibitemOpen
  \bibfield  {author} {\bibinfo {author} {\bibfnamefont {D.~A.}\ \bibnamefont
  {Lidar}},\ }\bibfield  {title} {\bibinfo {title} {Lecture notes on the theory
  of open quantum systems},\ }\href {http://arxiv.org/abs/1902.00967}
  {\bibfield  {journal} {\bibinfo  {journal} {arXiv preprint arXiv:1902.00967}\
  } (\bibinfo {year} {2019})}\BibitemShut {NoStop}%
\bibitem [{\citenamefont {Ciccarello}\ \emph {et~al.}(2022)\citenamefont
  {Ciccarello}, \citenamefont {Lorenzo}, \citenamefont {Giovannetti},\ and\
  \citenamefont {Palma}}]{Ciccarello2022Quantum}%
  \BibitemOpen
  \bibfield  {author} {\bibinfo {author} {\bibfnamefont {F.}~\bibnamefont
  {Ciccarello}}, \bibinfo {author} {\bibfnamefont {S.}~\bibnamefont {Lorenzo}},
  \bibinfo {author} {\bibfnamefont {V.}~\bibnamefont {Giovannetti}},\ and\
  \bibinfo {author} {\bibfnamefont {G.~M.}\ \bibnamefont {Palma}},\ }\bibfield
  {title} {\bibinfo {title} {{Quantum collision models: Open system dynamics
  from repeated interactions}},\ }\href
  {https://doi.org/10.1016/j.physrep.2022.01.001} {\bibfield  {journal}
  {\bibinfo  {journal} {Physics Reports}\ }\textbf {\bibinfo {volume} {954}},\
  \bibinfo {pages} {1} (\bibinfo {year} {2022})}\BibitemShut {NoStop}%
\bibitem [{\citenamefont {Leibfried}\ \emph {et~al.}(2003)\citenamefont
  {Leibfried}, \citenamefont {Blatt}, \citenamefont {Monroe},\ and\
  \citenamefont {Wineland}}]{Leibfried2003Quantum}%
  \BibitemOpen
  \bibfield  {author} {\bibinfo {author} {\bibfnamefont {D.}~\bibnamefont
  {Leibfried}}, \bibinfo {author} {\bibfnamefont {R.}~\bibnamefont {Blatt}},
  \bibinfo {author} {\bibfnamefont {C.}~\bibnamefont {Monroe}},\ and\ \bibinfo
  {author} {\bibfnamefont {D.}~\bibnamefont {Wineland}},\ }\bibfield  {title}
  {\bibinfo {title} {{Quantum dynamics of single trapped ions}},\ }\href
  {https://doi.org/10.1103/RevModPhys.75.281} {\bibfield  {journal} {\bibinfo
  {journal} {Reviews of Modern Physics}\ }\textbf {\bibinfo {volume} {75}},\
  \bibinfo {pages} {281} (\bibinfo {year} {2003})}\BibitemShut {NoStop}%
\bibitem [{\citenamefont {Rivas}\ \emph {et~al.}(2010)\citenamefont {Rivas},
  \citenamefont {{K Plato}}, \citenamefont {Huelga},\ and\ \citenamefont {{B
  Plenio}}}]{Rivas2010Markovian}%
  \BibitemOpen
  \bibfield  {author} {\bibinfo {author} {\bibfnamefont {{\'{A}}.}~\bibnamefont
  {Rivas}}, \bibinfo {author} {\bibfnamefont {A.~D.}\ \bibnamefont {{K
  Plato}}}, \bibinfo {author} {\bibfnamefont {S.~F.}\ \bibnamefont {Huelga}},\
  and\ \bibinfo {author} {\bibfnamefont {M.}~\bibnamefont {{B Plenio}}},\
  }\bibfield  {title} {\bibinfo {title} {{Markovian master equations: a
  critical study}},\ }\href {https://doi.org/10.1088/1367-2630/12/11/113032}
  {\bibfield  {journal} {\bibinfo  {journal} {New Journal of Physics}\ }\textbf
  {\bibinfo {volume} {12}},\ \bibinfo {pages} {113032} (\bibinfo {year}
  {2010})}\BibitemShut {NoStop}%
\bibitem [{\citenamefont {Breuer}\ and\ \citenamefont
  {Petruccione}(2007)}]{Breuer2007Theory}%
  \BibitemOpen
  \bibfield  {author} {\bibinfo {author} {\bibfnamefont {H.-P.}\ \bibnamefont
  {Breuer}}\ and\ \bibinfo {author} {\bibfnamefont {F.}~\bibnamefont
  {Petruccione}},\ }\href
  {https://doi.org/10.1093/acprof:oso/9780199213900.001.0001} {\emph {\bibinfo
  {title} {{The Theory of Open Quantum Systems}}}}\ (\bibinfo  {publisher}
  {Oxford University Press},\ \bibinfo {year} {2007})\BibitemShut {NoStop}%
\bibitem [{\citenamefont {Bhattacharyya}(1943)}]{Bhattacharyya1943Measure}%
  \BibitemOpen
  \bibfield  {author} {\bibinfo {author} {\bibfnamefont {A.}~\bibnamefont
  {Bhattacharyya}},\ }\bibfield  {title} {\bibinfo {title} {On a measure of
  divergence between two statistical populations defined by their probability
  distribution},\ }\href@noop {} {\bibfield  {journal} {\bibinfo  {journal}
  {Bulletin of the Calcutta Mathematical Society}\ }\textbf {\bibinfo {volume}
  {35}},\ \bibinfo {pages} {99} (\bibinfo {year} {1943})}\BibitemShut {NoStop}%
\bibitem [{\citenamefont {Goold}\ \emph {et~al.}(2016)\citenamefont {Goold},
  \citenamefont {Huber}, \citenamefont {Riera}, \citenamefont {del Rio},\ and\
  \citenamefont {Skrzypczyk}}]{Goold2016Role}%
  \BibitemOpen
  \bibfield  {author} {\bibinfo {author} {\bibfnamefont {J.}~\bibnamefont
  {Goold}}, \bibinfo {author} {\bibfnamefont {M.}~\bibnamefont {Huber}},
  \bibinfo {author} {\bibfnamefont {A.}~\bibnamefont {Riera}}, \bibinfo
  {author} {\bibfnamefont {L.}~\bibnamefont {del Rio}},\ and\ \bibinfo {author}
  {\bibfnamefont {P.}~\bibnamefont {Skrzypczyk}},\ }\bibfield  {title}
  {\bibinfo {title} {{The role of quantum information in thermodynamics—a
  topical review}},\ }\href {https://doi.org/10.1088/1751-8113/49/14/143001}
  {\bibfield  {journal} {\bibinfo  {journal} {Journal of Physics A:
  Mathematical and Theoretical}\ }\textbf {\bibinfo {volume} {49}},\ \bibinfo
  {pages} {143001} (\bibinfo {year} {2016})}\BibitemShut {NoStop}%
\bibitem [{\citenamefont {Scandi}\ and\ \citenamefont
  {Perarnau-Llobet}(2019)}]{Scandi2019Thermodynamic}%
  \BibitemOpen
  \bibfield  {author} {\bibinfo {author} {\bibfnamefont {M.}~\bibnamefont
  {Scandi}}\ and\ \bibinfo {author} {\bibfnamefont {M.}~\bibnamefont
  {Perarnau-Llobet}},\ }\bibfield  {title} {\bibinfo {title} {{Thermodynamic
  length in open quantum systems}},\ }\href
  {https://doi.org/10.22331/q-2019-10-24-197} {\bibfield  {journal} {\bibinfo
  {journal} {Quantum}\ }\textbf {\bibinfo {volume} {3}},\ \bibinfo {pages}
  {197} (\bibinfo {year} {2019})}\BibitemShut {NoStop}%
\bibitem [{\citenamefont {Miller}\ \emph {et~al.}(2019)\citenamefont {Miller},
  \citenamefont {Scandi}, \citenamefont {Anders},\ and\ \citenamefont
  {Perarnau-Llobet}}]{Miller2019Work}%
  \BibitemOpen
  \bibfield  {author} {\bibinfo {author} {\bibfnamefont {H.~J.~D.}\
  \bibnamefont {Miller}}, \bibinfo {author} {\bibfnamefont {M.}~\bibnamefont
  {Scandi}}, \bibinfo {author} {\bibfnamefont {J.}~\bibnamefont {Anders}},\
  and\ \bibinfo {author} {\bibfnamefont {M.}~\bibnamefont {Perarnau-Llobet}},\
  }\bibfield  {title} {\bibinfo {title} {{Work Fluctuations in Slow Processes:
  Quantum Signatures and Optimal Control}},\ }\href
  {https://doi.org/10.1103/PhysRevLett.123.230603} {\bibfield  {journal}
  {\bibinfo  {journal} {Physical Review Letters}\ }\textbf {\bibinfo {volume}
  {123}},\ \bibinfo {pages} {230603} (\bibinfo {year} {2019})}\BibitemShut
  {NoStop}%
\bibitem [{\citenamefont {Hermans}(1991)}]{Hermans1991Simple}%
  \BibitemOpen
  \bibfield  {author} {\bibinfo {author} {\bibfnamefont {J.}~\bibnamefont
  {Hermans}},\ }\bibfield  {title} {\bibinfo {title} {{Simple analysis of noise
  and hysteresis in (slow-growth) free energy simulations}},\ }\href
  {https://doi.org/10.1021/j100176a002} {\bibfield  {journal} {\bibinfo
  {journal} {The Journal of Physical Chemistry}\ }\textbf {\bibinfo {volume}
  {95}},\ \bibinfo {pages} {9029} (\bibinfo {year} {1991})}\BibitemShut
  {NoStop}%
\bibitem [{\citenamefont {Jarzynski}(1997)}]{Jarzynski1997Nonequilibrium}%
  \BibitemOpen
  \bibfield  {author} {\bibinfo {author} {\bibfnamefont {C.}~\bibnamefont
  {Jarzynski}},\ }\bibfield  {title} {\bibinfo {title} {{Nonequilibrium
  Equality for Free Energy Differences}},\ }\href
  {https://doi.org/10.1103/PhysRevLett.78.2690} {\bibfield  {journal} {\bibinfo
   {journal} {Physical Review Letters}\ }\textbf {\bibinfo {volume} {78}},\
  \bibinfo {pages} {2690} (\bibinfo {year} {1997})}\BibitemShut {NoStop}%
\bibitem [{\citenamefont {Petz}(2002)}]{Petz2002Covariance}%
  \BibitemOpen
  \bibfield  {author} {\bibinfo {author} {\bibfnamefont {D.}~\bibnamefont
  {Petz}},\ }\bibfield  {title} {\bibinfo {title} {{Covariance and Fisher
  information in quantum mechanics}},\ }\href
  {https://doi.org/10.1088/0305-4470/35/4/305} {\bibfield  {journal} {\bibinfo
  {journal} {Journal of Physics A: Mathematical and General}\ }\textbf
  {\bibinfo {volume} {35}},\ \bibinfo {pages} {929} (\bibinfo {year}
  {2002})}\BibitemShut {NoStop}%
\bibitem [{\citenamefont {Miller}\ and\ \citenamefont
  {Anders}(2018)}]{Miller2018Energy}%
  \BibitemOpen
  \bibfield  {author} {\bibinfo {author} {\bibfnamefont {H.~J.~D.}\
  \bibnamefont {Miller}}\ and\ \bibinfo {author} {\bibfnamefont
  {J.}~\bibnamefont {Anders}},\ }\bibfield  {title} {\bibinfo {title}
  {{Energy-temperature uncertainty relation in quantum thermodynamics}},\
  }\href {https://doi.org/10.1038/s41467-018-04536-7} {\bibfield  {journal}
  {\bibinfo  {journal} {Nature Communications}\ }\textbf {\bibinfo {volume}
  {9}},\ \bibinfo {pages} {2203} (\bibinfo {year} {2018})}\BibitemShut
  {NoStop}%
\bibitem [{\citenamefont {Luo}(2005)}]{Luo2005Quantum}%
  \BibitemOpen
  \bibfield  {author} {\bibinfo {author} {\bibfnamefont {S.~L.}\ \bibnamefont
  {Luo}},\ }\bibfield  {title} {\bibinfo {title} {{Quantum versus classical
  uncertainty}},\ }\href {https://doi.org/10.1007/s11232-005-0098-6} {\bibfield
   {journal} {\bibinfo  {journal} {Theoretical and Mathematical Physics}\
  }\textbf {\bibinfo {volume} {143}},\ \bibinfo {pages} {681} (\bibinfo {year}
  {2005})}\BibitemShut {NoStop}%
\bibitem [{\citenamefont {Gibilisco}\ \emph {et~al.}(2008)\citenamefont
  {Gibilisco}, \citenamefont {Imparato},\ and\ \citenamefont
  {Isola}}]{Gibilisco2009Inequalities}%
  \BibitemOpen
  \bibfield  {author} {\bibinfo {author} {\bibfnamefont {P.}~\bibnamefont
  {Gibilisco}}, \bibinfo {author} {\bibfnamefont {D.}~\bibnamefont
  {Imparato}},\ and\ \bibinfo {author} {\bibfnamefont {T.}~\bibnamefont
  {Isola}},\ }\bibfield  {title} {\bibinfo {title} {{Inequalities for quantum
  Fisher information}},\ }\href {https://doi.org/10.1090/S0002-9939-08-09447-1}
  {\bibfield  {journal} {\bibinfo  {journal} {Proceedings of the American
  Mathematical Society}\ }\textbf {\bibinfo {volume} {137}},\ \bibinfo {pages}
  {317} (\bibinfo {year} {2008})}\BibitemShut {NoStop}%
\bibitem [{\citenamefont {Fr{\'{e}}rot}\ and\ \citenamefont
  {Roscilde}(2016)}]{Frerot2016Quantum}%
  \BibitemOpen
  \bibfield  {author} {\bibinfo {author} {\bibfnamefont {I.}~\bibnamefont
  {Fr{\'{e}}rot}}\ and\ \bibinfo {author} {\bibfnamefont {T.}~\bibnamefont
  {Roscilde}},\ }\bibfield  {title} {\bibinfo {title} {{Quantum variance: A
  measure of quantum coherence and quantum correlations for many-body
  systems}},\ }\href {https://doi.org/10.1103/PhysRevB.94.075121} {\bibfield
  {journal} {\bibinfo  {journal} {Physical Review B}\ }\textbf {\bibinfo
  {volume} {94}},\ \bibinfo {pages} {075121} (\bibinfo {year}
  {2016})}\BibitemShut {NoStop}%
\bibitem [{\citenamefont {Konopik}\ and\ \citenamefont
  {Lutz}(2019)}]{Konopik2019Quantum}%
  \BibitemOpen
  \bibfield  {author} {\bibinfo {author} {\bibfnamefont {M.}~\bibnamefont
  {Konopik}}\ and\ \bibinfo {author} {\bibfnamefont {E.}~\bibnamefont {Lutz}},\
  }\bibfield  {title} {\bibinfo {title} {{Quantum response theory for
  nonequilibrium steady states}},\ }\href
  {https://doi.org/10.1103/PhysRevResearch.1.033156} {\bibfield  {journal}
  {\bibinfo  {journal} {Physical Review Research}\ }\textbf {\bibinfo {volume}
  {1}},\ \bibinfo {pages} {033156} (\bibinfo {year} {2019})}\BibitemShut
  {NoStop}%
\bibitem [{\citenamefont {Levy}\ \emph {et~al.}(2021)\citenamefont {Levy},
  \citenamefont {Rabani},\ and\ \citenamefont {Limmer}}]{Levy2021Response}%
  \BibitemOpen
  \bibfield  {author} {\bibinfo {author} {\bibfnamefont {A.}~\bibnamefont
  {Levy}}, \bibinfo {author} {\bibfnamefont {E.}~\bibnamefont {Rabani}},\ and\
  \bibinfo {author} {\bibfnamefont {D.~T.}\ \bibnamefont {Limmer}},\ }\bibfield
   {title} {\bibinfo {title} {{Response theory for nonequilibrium steady states
  of open quantum systems}},\ }\href
  {https://doi.org/10.1103/PhysRevResearch.3.023252} {\bibfield  {journal}
  {\bibinfo  {journal} {Physical Review Research}\ }\textbf {\bibinfo {volume}
  {3}},\ \bibinfo {pages} {023252} (\bibinfo {year} {2021})}\BibitemShut
  {NoStop}%
\bibitem [{\citenamefont {Gibilisco}\ and\ \citenamefont
  {Isola}(2003)}]{Gibilisco2003Wigner}%
  \BibitemOpen
  \bibfield  {author} {\bibinfo {author} {\bibfnamefont {P.}~\bibnamefont
  {Gibilisco}}\ and\ \bibinfo {author} {\bibfnamefont {T.}~\bibnamefont
  {Isola}},\ }\bibfield  {title} {\bibinfo {title} {{Wigner–Yanase
  information on quantum state space: The geometric approach}},\ }\href
  {https://doi.org/10.1063/1.1598279} {\bibfield  {journal} {\bibinfo
  {journal} {Journal of Mathematical Physics}\ }\textbf {\bibinfo {volume}
  {44}},\ \bibinfo {pages} {3752} (\bibinfo {year} {2003})}\BibitemShut
  {NoStop}%
\bibitem [{\citenamefont {Albash}\ \emph {et~al.}(2012)\citenamefont {Albash},
  \citenamefont {Boixo}, \citenamefont {Lidar},\ and\ \citenamefont
  {Zanardi}}]{Albash2012Quantum}%
  \BibitemOpen
  \bibfield  {author} {\bibinfo {author} {\bibfnamefont {T.}~\bibnamefont
  {Albash}}, \bibinfo {author} {\bibfnamefont {S.}~\bibnamefont {Boixo}},
  \bibinfo {author} {\bibfnamefont {D.~A.}\ \bibnamefont {Lidar}},\ and\
  \bibinfo {author} {\bibfnamefont {P.}~\bibnamefont {Zanardi}},\ }\bibfield
  {title} {\bibinfo {title} {{Quantum adiabatic Markovian master equations}},\
  }\href {https://doi.org/10.1088/1367-2630/14/12/123016} {\bibfield  {journal}
  {\bibinfo  {journal} {New Journal of Physics}\ }\textbf {\bibinfo {volume}
  {14}},\ \bibinfo {pages} {123016} (\bibinfo {year} {2012})}\BibitemShut
  {NoStop}%
\bibitem [{\citenamefont {Horowitz}\ and\ \citenamefont
  {Gingrich}(2020)}]{Horowitz2020Thermodynamic}%
  \BibitemOpen
  \bibfield  {author} {\bibinfo {author} {\bibfnamefont {J.~M.}\ \bibnamefont
  {Horowitz}}\ and\ \bibinfo {author} {\bibfnamefont {T.~R.}\ \bibnamefont
  {Gingrich}},\ }\bibfield  {title} {\bibinfo {title} {{Thermodynamic
  uncertainty relations constrain non-equilibrium fluctuations}},\ }\href
  {https://doi.org/10.1038/s41567-019-0702-6} {\bibfield  {journal} {\bibinfo
  {journal} {Nature Physics}\ }\textbf {\bibinfo {volume} {16}},\ \bibinfo
  {pages} {15} (\bibinfo {year} {2020})}\BibitemShut {NoStop}%
\bibitem [{\citenamefont {Gyhm}\ \emph {et~al.}(2022)\citenamefont {Gyhm},
  \citenamefont {{\v{S}}afr{\'{a}}nek},\ and\ \citenamefont
  {Rosa}}]{Gyhm2022Quantum}%
  \BibitemOpen
  \bibfield  {author} {\bibinfo {author} {\bibfnamefont {J.-Y.}\ \bibnamefont
  {Gyhm}}, \bibinfo {author} {\bibfnamefont {D.}~\bibnamefont
  {{\v{S}}afr{\'{a}}nek}},\ and\ \bibinfo {author} {\bibfnamefont
  {D.}~\bibnamefont {Rosa}},\ }\bibfield  {title} {\bibinfo {title} {{Quantum
  Charging Advantage Cannot Be Extensive without Global Operations}},\ }\href
  {https://doi.org/10.1103/PhysRevLett.128.140501} {\bibfield  {journal}
  {\bibinfo  {journal} {Physical Review Letters}\ }\textbf {\bibinfo {volume}
  {128}},\ \bibinfo {pages} {140501} (\bibinfo {year} {2022})}\BibitemShut
  {NoStop}%
\bibitem [{\citenamefont {Mohan}\ \emph {et~al.}(2022)\citenamefont {Mohan},
  \citenamefont {Das},\ and\ \citenamefont {Pati}}]{Mohan2022Quantum}%
  \BibitemOpen
  \bibfield  {author} {\bibinfo {author} {\bibfnamefont {B.}~\bibnamefont
  {Mohan}}, \bibinfo {author} {\bibfnamefont {S.}~\bibnamefont {Das}},\ and\
  \bibinfo {author} {\bibfnamefont {A.~K.}\ \bibnamefont {Pati}},\ }\bibfield
  {title} {\bibinfo {title} {{Quantum speed limits for information and
  coherence}},\ }\href {https://doi.org/10.1088/1367-2630/ac753c} {\bibfield
  {journal} {\bibinfo  {journal} {New Journal of Physics}\ }\textbf {\bibinfo
  {volume} {24}},\ \bibinfo {pages} {065003} (\bibinfo {year}
  {2022})}\BibitemShut {NoStop}%
\bibitem [{\citenamefont {H{\"{u}}bner}(1992)}]{Hubner1992Explicit}%
  \BibitemOpen
  \bibfield  {author} {\bibinfo {author} {\bibfnamefont {M.}~\bibnamefont
  {H{\"{u}}bner}},\ }\bibfield  {title} {\bibinfo {title} {{Explicit
  computation of the Bures distance for density matrices}},\ }\href
  {https://doi.org/10.1016/0375-9601(92)91004-B} {\bibfield  {journal}
  {\bibinfo  {journal} {Physics Letters A}\ }\textbf {\bibinfo {volume}
  {163}},\ \bibinfo {pages} {239} (\bibinfo {year} {1992})}\BibitemShut
  {NoStop}%
\bibitem [{\citenamefont {Augusiak}\ \emph {et~al.}(2016)\citenamefont
  {Augusiak}, \citenamefont {Ko{\l}ody{\'{n}}ski}, \citenamefont {Streltsov},
  \citenamefont {Bera}, \citenamefont {Ac{\'{i}}n},\ and\ \citenamefont
  {Lewenstein}}]{Augusiak2016Asymptotic}%
  \BibitemOpen
  \bibfield  {author} {\bibinfo {author} {\bibfnamefont {R.}~\bibnamefont
  {Augusiak}}, \bibinfo {author} {\bibfnamefont {J.}~\bibnamefont
  {Ko{\l}ody{\'{n}}ski}}, \bibinfo {author} {\bibfnamefont {A.}~\bibnamefont
  {Streltsov}}, \bibinfo {author} {\bibfnamefont {M.~N.}\ \bibnamefont {Bera}},
  \bibinfo {author} {\bibfnamefont {A.}~\bibnamefont {Ac{\'{i}}n}},\ and\
  \bibinfo {author} {\bibfnamefont {M.}~\bibnamefont {Lewenstein}},\ }\bibfield
   {title} {\bibinfo {title} {{Asymptotic role of entanglement in quantum
  metrology}},\ }\href {https://doi.org/10.1103/PhysRevA.94.012339} {\bibfield
  {journal} {\bibinfo  {journal} {Physical Review A}\ }\textbf {\bibinfo
  {volume} {94}},\ \bibinfo {pages} {012339} (\bibinfo {year}
  {2016})}\BibitemShut {NoStop}%
\bibitem [{\citenamefont {Horn}\ and\ \citenamefont
  {Johnson}(1985)}]{Horn1985Matrix}%
  \BibitemOpen
  \bibfield  {author} {\bibinfo {author} {\bibfnamefont {R.~A.}\ \bibnamefont
  {Horn}}\ and\ \bibinfo {author} {\bibfnamefont {C.~R.}\ \bibnamefont
  {Johnson}},\ }\href {https://doi.org/10.1017/CBO9780511810817} {\emph
  {\bibinfo {title} {{Matrix analysis}}}},\ Vol.~\bibinfo {volume} {72}\
  (\bibinfo  {publisher} {Cambridge University Press},\ \bibinfo {address}
  {Cambridge},\ \bibinfo {year} {1985})\ pp.\ \bibinfo {pages}
  {692--692}\BibitemShut {NoStop}%
\bibitem [{\citenamefont {Bu{\v{c}}a}\ and\ \citenamefont
  {Prosen}(2012)}]{Buca2012Note}%
  \BibitemOpen
  \bibfield  {author} {\bibinfo {author} {\bibfnamefont {B.}~\bibnamefont
  {Bu{\v{c}}a}}\ and\ \bibinfo {author} {\bibfnamefont {T.}~\bibnamefont
  {Prosen}},\ }\bibfield  {title} {\bibinfo {title} {{A note on symmetry
  reductions of the Lindblad equation: transport in constrained open spin
  chains}},\ }\href {https://doi.org/10.1088/1367-2630/14/7/073007} {\bibfield
  {journal} {\bibinfo  {journal} {New Journal of Physics}\ }\textbf {\bibinfo
  {volume} {14}},\ \bibinfo {pages} {073007} (\bibinfo {year}
  {2012})}\BibitemShut {NoStop}%
\bibitem [{\citenamefont {Sakurai}\ and\ \citenamefont
  {Commins}(1995)}]{sakurai1995modern}%
  \BibitemOpen
  \bibfield  {author} {\bibinfo {author} {\bibfnamefont {J.~J.}\ \bibnamefont
  {Sakurai}}\ and\ \bibinfo {author} {\bibfnamefont {E.~D.}\ \bibnamefont
  {Commins}},\ }\href@noop {} {\bibinfo {title} {Modern quantum mechanics,
  revised edition}} (\bibinfo {year} {1995})\BibitemShut {NoStop}%
\bibitem [{\citenamefont {Kempe}\ \emph {et~al.}(2001)\citenamefont {Kempe},
  \citenamefont {Bacon}, \citenamefont {Lidar},\ and\ \citenamefont
  {Whaley}}]{Kempe2001Theory}%
  \BibitemOpen
  \bibfield  {author} {\bibinfo {author} {\bibfnamefont {J.}~\bibnamefont
  {Kempe}}, \bibinfo {author} {\bibfnamefont {D.}~\bibnamefont {Bacon}},
  \bibinfo {author} {\bibfnamefont {D.~A.}\ \bibnamefont {Lidar}},\ and\
  \bibinfo {author} {\bibfnamefont {K.~B.}\ \bibnamefont {Whaley}},\ }\bibfield
   {title} {\bibinfo {title} {{Theory of decoherence-free fault-tolerant
  universal quantum computation}},\ }\href
  {https://doi.org/10.1103/PhysRevA.63.042307} {\bibfield  {journal} {\bibinfo
  {journal} {Physical Review A}\ }\textbf {\bibinfo {volume} {63}},\ \bibinfo
  {pages} {042307} (\bibinfo {year} {2001})}\BibitemShut {NoStop}%
\bibitem [{\citenamefont {Kubo}(1957)}]{Kubo1957Statistical}%
  \BibitemOpen
  \bibfield  {author} {\bibinfo {author} {\bibfnamefont {R.}~\bibnamefont
  {Kubo}},\ }\bibfield  {title} {\bibinfo {title} {{Statistical-Mechanical
  Theory of Irreversible Processes. I. General Theory and Simple Applications
  to Magnetic and Conduction Problems}},\ }\href
  {https://doi.org/10.1143/JPSJ.12.570} {\bibfield  {journal} {\bibinfo
  {journal} {Journal of the Physical Society of Japan}\ }\textbf {\bibinfo
  {volume} {12}},\ \bibinfo {pages} {570} (\bibinfo {year} {1957})}\BibitemShut
  {NoStop}%
\bibitem [{\citenamefont {Hansen}(2008)}]{Hansen2008Metric}%
  \BibitemOpen
  \bibfield  {author} {\bibinfo {author} {\bibfnamefont {F.}~\bibnamefont
  {Hansen}},\ }\bibfield  {title} {\bibinfo {title} {{Metric adjusted skew
  information}},\ }\href {https://doi.org/10.1073/pnas.0803323105} {\bibfield
  {journal} {\bibinfo  {journal} {Proceedings of the National Academy of
  Sciences}\ }\textbf {\bibinfo {volume} {105}},\ \bibinfo {pages} {9909}
  (\bibinfo {year} {2008})}\BibitemShut {NoStop}%
\end{thebibliography}%

\onecolumngrid
\appendix

\section{Derivation of Result 1} \label{app:main_derivation}
Here, we present a detailed derivation of Result 1 in the main text.
We begin by time-evolving the states of the two trajectories at time $t$ for a short additional time $\delta t$:
    \begin{align} \label{eqn:rhosigma}
        \rho_{t+\delta t}
        &= \rho_t + \delta t \mc{L}_t(\rho_t) + \mc{O}(\delta t^2),
        \\
        \sigma_{t+\delta t}
        &= \sigma_t + \delta t \mc{L}_t(\sigma_t) + \delta t \mc{P}_t(\sigma_t) + \mc{O}(\delta t^2).
    \end{align}
    We also denote by $\sigma'_{t+\delta t}$ the state evolved $\sigma_t$ under the unperturbed dynamics for time $\delta t$, as illustrated in Fig.~\ref{fig:trajectories} in the main text:
    \begin{equation} \label{eqn:sigmaprime}
        \sigma'_{t+\delta t} = \sigma_t + \delta t \mc{L}_t(\sigma_t) + \mc{O}(\delta t^2).
    \end{equation}
    Then we obtain the following:
    \begin{align} 
        \bures(\rho_{t+\delta t}, \sigma_{t+\delta t}) \nonumber
            &\leq
            \bures(\rho_{t+\delta t}, \sigma'_{t+\delta t}) + \bures(\sigma'_{t+\delta t}, \sigma_{t + \delta t}) \nonumber \\
            &= \bures(\mc{N}_{t+\delta t,t}(\rho_t), \mc{N}_{t+\delta t,t}(\sigma_t)) + \bures(\sigma'_{t+\delta t}, \sigma_{t + \delta t}) \nonumber \\
            &\leq \bures(\rho_t, \sigma_t) + \bures(\sigma'_{t+\delta t}, \sigma_{t + \delta t}).
    \end{align}
    In the first line, we use the fact that the Bures angle obeys the triangle inequality, $\bures(\rho_1, \rho_3) \leq \bures(\rho_1, \rho_2) + \bures(\rho_2, \rho_3)$ for any quantum states $\rho_1, \rho_2, \rho_3$~\cite{NielsenChuang_ch9}.
    In the second line, we use the divisibility of the Markovian GKSL dynamics, implying that the mapping from time $t$ to time $t + \delta t$ is described by a quantum channel $\mc{N}_{t+\delta t, t}$.
    The last line follows from the contractivity of the Bures angle under quantum channels: $\bures(\mathcal{E}(\rho_1), \mathcal{E}(\rho_2)) \leq \bures(\rho_1, \rho_2)$ for any states $\rho_1, \rho_2$ and any channel $\mc{E}$.
    Rearranging, we therefore have
    \begin{align} \label{eqn:res1proof_intermediate}
        \bures(\rho_{t+\delta t},\sigma_{t+\delta t}) - \bures(\rho_t, \sigma_t) \leq \bures(\sigma'_{t+\delta t}, \sigma_{t+\delta t}).
    \end{align}
    For the right-hand side of Eq.~\eqref{eqn:res1proof_intermediate}, we use the infinitesimal form of the Bures angle:
    $\bures(\rho, \rho + \dd \rho)^2 = \frac{1}{2} \sum_{i,j} \frac{\abs{ \mel{\psi_i}{\dd \rho}{\psi_j} }^2}{\lambda_i+ \lambda_j}$
    for the spectral decomposition $\rho = \sum_i \lambda_i \ket{\psi_i}\!\bra{\psi_i}$ with eigenvalues $\lambda_i$ and corresponding eigenvectors $\ket{\psi_i}$, and an infinitesimal translation $\dd \rho$~\cite{Hubner1992Explicit}.
    Here, we replace $\rho \to \sigma_{t+\delta t}$ and use the small finite difference $\dd \rho \to \sigma'_{t+\delta t} - \sigma_{t+ \delta t} = \delta t \mc{P}_t(\sigma_t) + \mc{O}(\delta t^2)$, from Eqs.~\eqref{eqn:rhosigma} and \eqref{eqn:sigmaprime}.
    Note that the infinitesimal expansion of the Bures angle agrees with the QFI expression given in the main text, so
    \begin{align}
        \bures(\sigma'_{t+\delta t}, \sigma_{t + \delta t})^2 = \frac{\delta t^2}{4} \mc{F}(\sigma_{t+\delta t}, \mc{P}_t) + \mc{O}(\delta t^3).
    \end{align}
    Inserting this expression into Eq.~\eqref{eqn:res1proof_intermediate} gives
    \begin{align}
        \bures(\rho_{t+\delta t},\sigma_{t+\delta t}) - \bures(\rho_t, \sigma_t) \leq \frac{\delta t}{2} \sqrt{\mc{F}(\sigma_{t+\delta t},\mc{P}_t)} + \mc{O}(\delta t^2),
    \end{align}
    and dividing by $\delta t$,
    \begin{align}
        \frac{\bures(\rho_{t+\delta t},\sigma_{t+\delta t}) - \bures(\rho_t, \sigma_t)}{\delta t} \leq \frac{1}{2} \sqrt{\mc{F}(\sigma_{t+\delta t},\mc{P}_t)} + \mc{O}(\delta t).
    \end{align}
    Letting $\delta t \to 0$, the left-hand side becomes a derivative, and the QFI on the right-hand side tends to $\mc{F}(\sigma_t, \mc{P}_t)$ by continuity~\cite{Augusiak2016Asymptotic}, hence
    \begin{align}
        \frac{\dd \bures(\rho_t, \sigma_t)}{\dd t} \leq \frac{1}{2} \sqrt{\mc{F}(\sigma_t, \mc{P}_t)}.
    \end{align}
    Finally, integrating from over times $s = 0$ to $t$ using the initial condition $\rho_0 = \sigma_0$ results in
    \begin{align}
        \bures(\rho_t, \sigma_t) \leq \frac{1}{2} \int_0^t \dd s \; \sqrt{\mc{F}(\sigma_s, \mc{P}_s)}.
    \end{align}
    This concludes the proof of Result 1.\\

    Here we also present a useful form for the right-hand side of the bound in the case of a fixed generator $\mc{V}$ multiplied by a time-dependent coefficient $v_t$.
    Applying the Cauchy-Schwarz inequality to the right-hand side of Eq.~\eqref{eqn:main_bound}, we have
    \begin{align}
        \int_0^t \dd s \; \sqrt{\mc{F}(\sigma_s, v_s \mc{V})} & = \int_0^t \dd s \; \abs{v_s} \sqrt{\mc{F}(\sigma_s, \mc{V})} \nonumber \\
            & \leq \left( \int_0^t \dd s \; v_s^2 \right)^{\frac{1}{2}} \left( \int_0^t \dd s \; \mc{F}(\sigma_s, \mc{V}) \right)^{\frac{1}{2}} \nonumber \\
            & = t \ev{v_s^2}_t^{\frac{1}{2}} \ev{\mc{F}(\sigma_s, \mc{V})}_t^{\frac{1}{2}},
    \end{align}
    where angled brackets $\ev{\cdot}_t$ denote a time average over $s \in [0,t]$.
    So the witness criterion Eq.~\eqref{eqn:witness} (main text) is modified by replacing $v$ with the root-mean-square $\ev{v_s^2}_t^{\frac{1}{2}}$.

    \section{Observable speed limit} \label{app:obs_speedlim}

    Consider an arbitrary observable $A$; we will bound the rate of change of the difference in expectation of $A$ between the two trajectories $\rho_t$ and $\sigma_t$.
    Following the same method as in the proof of Result 1, we start by using Eqs.~\eqref{eqn:rhosigma} and \eqref{eqn:sigmaprime} to write
    \begin{align}
        \rho_{t+\delta t} - \sigma_{t+\delta t} & = (\rho_{t+\delta t} - \sigma'_{t+\delta t}) + (\sigma'_{t+\delta t} - \sigma_{t+\delta t}) \nonumber \\
        & = \rho_t - \sigma_t + \delta t [\mc{L}_t(\rho_t) - \mc{L}_t(\sigma_t)] + \delta t \mc{P}_t(\sigma_t) + \mc{O}(\delta t^2),
    \end{align}
    therefore
    \begin{align}
        \tr[A(\rho_{t+\delta t} - \sigma_{t+\delta t})] - \tr[A(\rho_t - \sigma_t)] & = \delta t \tr[A \mc{L}_t(\rho_t - \sigma_t)] + \delta t \tr[A \mc{P}_t(\sigma_t)] + \mc{O}(\delta t^2).
    \end{align}
    Dividing by $\delta t$ and using the triangle inequality, we have
    \begin{align}
        \frac{1}{\delta t} \abs{\tr[A(\rho_{t+\delta t} - \sigma_{t+\delta t})] - \tr[A(\rho_t - \sigma_t)] } & \leq \abs{\tr[A\mc{L}_t(\rho_t - \sigma_t)]} + \abs{\tr[A\mc{P}_t(\sigma_t)]} + \mc{O}(\delta t),
    \end{align}
    and taking $\delta t \to 0$ implies
    \begin{align} \label{eqn:obs_speed_intermediate}
        \abs{ \frac{\dd}{\dd t} \tr[A(\rho_t - \sigma_t)] } & \leq \abs{\tr[A\mc{L}_t(\rho_t - \sigma_t)]} + \abs{\tr[A\mc{P}_t(\sigma_t)]}.
    \end{align}
    For the first term on the right-hand side of Eq.~\eqref{eqn:obs_speed_intermediate}, we use the adjoint generator $\mc{L}_t^\dagger$ -- defined by $\tr[X \mc{L}_t(Y)] = \tr[\mc{L}_t^\dagger(X) Y]$ for all operators $X,Y$ -- to give
    \begin{align}
        \abs{\tr[A\mc{L}_t(\rho_t - \sigma_t)]} & = \abs{\tr[\mc{L}_t^\dagger(A) (\rho_t-\sigma_t) ]} \nonumber \\
            & \leq \norm*{\mc{L}_t^\dagger(A)} \|\rho_t - \sigma_t\|_1 \nonumber \\
            & = 2\norm*{\mc{L}_t^\dagger(A)} \, D_{\tr}(\rho_t,\sigma_t),
    \end{align}
    having introduced the trace distance $D_{\tr}(\rho,\sigma) := \frac{1}{2}\|\rho - \sigma\|_1$ and used H\"older's inequality~\cite[Appendix B]{Horn1985Matrix}.\\
    
    The remaining term in Eq.~\eqref{eqn:obs_speed_intermediate} is dealt with in the same way as in derivations of other observable speed limits~\cite{GarciaPintos2022Unifying}.
    This employs the symmetric logarithmic derivative (SLD) $L_t$, here defined implicitly by the relation
    \begin{align}
        \frac{1}{2} \{ L_t, \sigma_t \} = \mc{P}_t(\sigma_t).
    \end{align}
    The SLD has vanishing mean, $\tr[L_t \sigma_t] = 0$, and variance equal to the QFI, $\tr[L_t^2 \sigma_t] = \mc{F}(\sigma_t, \mc{P}_t)$.
    The cyclic property of the trace implies
    \begin{align}
        \tr[A \mc{P}_t(\sigma_t)] & = \frac{1}{2} \tr[A \{ L_t, \sigma_t \}] \nonumber \\
            & = \frac{1}{2} \tr[\sigma_t \{L_t, A \}].
    \end{align}
    This quantity can be recognised as the covariance of $L_t$ and $A$ in the state $\sigma_t$; the Cauchy-Schwarz inequality then implies
    \begin{align}
        \abs{\tr[A \mc{P}_t(\sigma_t)]} & \leq \sqrt{\variance(\sigma_t, A) \variance(\sigma_t, L_t)} \nonumber \\
            & = \sqrt{\variance(\sigma_t, A) \mc{F}(\sigma_t, \mc{P}_t)}.
    \end{align}
    Putting everything together, we therefore have the observable speed limit (in terms of the derivative)
    \begin{align} \label{eqn:obs_speed_lim}
        \abs{ \frac{\dd}{\dd t} \tr[A(\rho_t - \sigma_t)] } & \leq 2\norm*{\mc{L}_t^\dagger(A)} \, D_{\tr}(\rho_t,\sigma_t) + \sqrt{\variance(\sigma_t, A) \mc{F}(\sigma_t, \mc{P}_t)}.
    \end{align}\\

    An upper bound on the trace distance appearing on the right-hand side can also be obtained using Result 1, since $D_{\tr} \leq \bures$~\cite{NielsenChuang_ch9}:
    \begin{align}
        D_{\tr}(\rho_t, \sigma_t) & \leq \frac{1}{2} \int_0^t \dd s \; \sqrt{\mc{F}(\sigma_s, \mc{P}_s)}.
    \end{align}
    
    A special case of interest is when $A$ is conserved by the unperturbed dynamics, for example when the master equation has a so-called strong symmetry~\cite{Buca2012Note}.
    If the perturbation breaks this symmetry, we can measure how much this is evidenced in the system's response via the change in expectation value of $A$.
    In this case, the first term in Eq.~\eqref{eqn:obs_speed_lim} vanishes because $\mc{L}_t^\dagger(A) = 0$.

\section{Weak coupling perturbation} \label{app:weak}
Here, we show the conditions under which the change to a perturbed open system's dynamics can be approximated as just containing the perturbing Hamiltonian $vV$, and estimate the size of the error.\\

First, we give details about the GKSL master equation.
We assume a standard weak-coupling master equation with secular approximation~\cite{Breuer2007Theory}, 
\begin{align}
    \frac{\dd \rho_t}{\dd t} = \mc{L}(\rho_t) = -i[H + H_\mathrm{LS}, \rho_t] + \mc{D}(\rho_t),
\end{align}
where the Lamb shift Hamiltonian $H_\mathrm{LS}$ and dissipator $\mc{D}$ are given by
\begin{align}
    H_\mathrm{LS} & = \lambda^2 \sum_{\omega,\alpha,\beta} S_{\alpha \beta}(\omega) A_\alpha^\dagger(\omega) A_\beta(\omega), \\
\mc{D}(\rho) & = \lambda ^2 \sum_{\omega, \alpha,\beta} \gamma_{\alpha \beta}(\omega) \Big[  A_\beta(\omega) \rho A_\alpha^\dagger(\omega) 
        - \frac{1}{2}\left\{ A_\alpha^\dagger(\omega) A_\beta(\omega), \rho \right\} \Big].
\end{align}
Here, the system-bath interaction $H_I = \lambda \sum_\alpha A_\alpha \otimes B_\alpha$ is decomposed into components $A_\alpha = \sum_\omega A_\alpha(\omega)$ with Bohr frequencies $\omega$ (i.e., gaps in the spectrum of the system Hamiltonian $H$), such that $[H, A_\alpha(\omega)] = \omega A_\alpha(\omega)$.
The coefficients are real and imaginary parts of the Fourier-transformed (stationary) bath correlation functions
\begin{align}
    \Gamma_{\alpha \beta}(\omega) & := \int_0^\infty \dd s \; e^{i \omega s} \ev{B_\alpha^\dagger(s) B_\beta(0)} \\
    & =: \frac{1}{2} \gamma_{\alpha \beta}(\omega) + i S_{\alpha \beta}(\omega),
\end{align}    
where $B_\alpha(s) = e^{is H_B}B_\alpha e^{-is H_B}$ is the bath operator $B_\alpha$ in the interaction picture, using the bath Hamiltonian $H_B$, and angled brackets denote expectation value.
We factor out the coupling strength $\lambda$ such that $A_\alpha,\, B_\alpha = \mc{O}(1)$ (independent of $\lambda$).\\

Recall the relevant timescales: $\tau_S \sim h^{-1}$ for the system Hamiltonian, $\tau_V \sim v^{-1}$ for the perturbation, $\tau_R \sim \lambda^{-2} \gamma^{-1}$ for the system relaxation, and $\tau_B$ for the bath correlation decay.
In terms of these, the assumptions being used are:
\begin{enumerate}
    \item[i)] Born-Markov approximation: $\tau_B \ll \tau_R$,
    \item[ii)] rotating wave approximation (RWA): $\tau_S \ll \tau_R$,
    \item[iii)] small perturbation relative to bath: $\tau_B \ll \tau_V$,
    \item[iv)] small perturbation relative to system: $\tau_S \ll \tau_V$.
\end{enumerate}
The need for assumption (iii) is not obvious but will become apparent later -- it is needed essentially so we can approximate the bath correlation function as flat in the frequency domain.
Firstly, we expand the perturbed dissipator $\mc{D}'$ and Lamb shift $H'_\mathrm{LS}$ to first order in $v$ using standard perturbation theory~\cite{sakurai1995modern}.
Note that we assume both $H = \sum_n E_n \dyad{n}$ and $H' = \sum_n E'_n \dyad{n'}$ to be non-degenerate.
To lowest order, the perturbed energy eigenvalues and eigenvectors are $E'_n = E_n + v E_n^{(1)} + \mc{O}(v^2)$, $\ket{n'} = \ket{n} + v \ket{n^{(1)}} + \mc{O}(v^2)$, with
\begin{align}
    E^{(1)}_n & = \mel{n}{V}{n}, \nonumber \\
    \ket*{n^{(1)}} & = \sum_{m \neq n} \frac{\mel{m}{V}{n}}{E_n-E_m} \ket{m} =: \sum_{m \neq n} C_{n m} \ket{m}.
\end{align}

The perturbed Bohr frequencies are of the form $\omega' \approx \omega + \delta \omega$ with $\delta \omega = \mc{O}(v)$.
It is possible for some $\omega$ to be degenerate, meaning that several transitions may have the same gap (for example, with the energies $\{-1,0,1\}$).
We initially assume that the pattern of Bohr frequency degeneracies is unchanged under the perturbation, showing below how to handle a breaking of degeneracy.
Recall that the RWA neglects terms in the master equation where $A_\alpha(\omega_1)$ and $A_\beta^\dagger(\omega_2)$ occur with $\abs{\omega_1 - \omega_2} \gg \tau_R^{-1}$~\cite{Breuer2007Theory}.
On the other hand, if $\abs{\omega_1 - \omega_2} \ll \tau_R^{-1}$, then we should treat these frequencies as effectively degenerate so that the RWA does not remove such off-diagonal terms, and we can include $A_\alpha(\omega_1)$ and $A_\alpha(\omega_2)$ in the same frequency component of $A_\alpha$.

Supposing that the perturbed frequencies $\omega'$ have the same pattern of degeneracies as the $\omega$, the perturbed jump operators are
\begin{align}
    A'_\alpha(\omega') & = \sum_{m,n : \: E'_m - E'_n \approx \omega'} \Pi'_m A_\alpha \Pi'_n \nonumber \\
        & = \sum_{m,n: \: E_m - E_n \approx \omega} \Pi'_m A_\alpha \Pi'_n \nonumber \\
        & = \sum_{m,n:\; E_m - E_n \approx \omega} \Pi_m A_\alpha \Pi_n + v\left( Q_m A_\alpha \Pi_n + \Pi_m A_\alpha Q_n \right)  + \mc{O}(v^2) \nonumber \\
        & =: A_\alpha(\omega) + v A^{(1)}_\alpha(\omega) + \mc{O}(v^2),
\end{align}
where the $\approx$ sign is shorthand for equality up to an error much less than $\tau_R^{-1}$, i.e., $\abs{E_m-E_n - \omega} \ll \tau_R^{-1}$, and
\begin{align}
    \Pi'_n & = \Pi_n + v Q_n + \mc{O}(v^2), \nonumber \\
    Q_n & := \sum_{m \neq n} C_{n m} \ketbra{m}{n} + C_{n m}^* \ketbra{n}{m}.
\end{align}
For the bath correlation coefficients, we write $\gamma_{\alpha \beta}(\omega') = \gamma_{\alpha \beta}(\omega) + \delta \omega \partial_\omega \gamma_{\alpha \beta}(\omega) + \mc{O}(v^2)$ and similarly for $S_{\alpha \beta}(\omega')$.
Inserting these into the expressions for $H_\mathrm{LS}$ and $\mc{D}$
to determine the perturbed Lamb shift and dissipator, we have the first order terms
\begin{align}
    v H^{(1)}_\mathrm{LS} & = \lambda^2 \sum_\omega \sum_{\alpha,\beta} v S_{\alpha \beta}(\omega) \left[ {A_\alpha^{(1)}}^\dagger A_\beta(\omega) + A_\alpha^\dagger(\omega) A_\beta^{(1)}(\omega) \right] + \delta \omega \partial_\omega S_{\alpha \beta}(\omega) A_\alpha^\dagger(\omega) A_\beta(\omega), \nonumber \\
    v \mc{D}^{(1)}(\rho) & = \lambda^2 \sum_\omega \sum_{\alpha,\beta} v \gamma_{\alpha \beta}(\omega) \left[ A_\beta^{(1)}(\omega)\rho A_\alpha^\dagger(\omega) + A_\beta(\omega) \rho {A_\alpha^{(1)}}^\dagger(\omega) - \frac{1}{2} \left\{ {A_\alpha^{(1)}}^\dagger(\omega) A_\beta(\omega) + A_\alpha^\dagger(\omega) A_\beta^{(1)}(\omega), \rho \right\} \right] \nonumber \\
        &  \quad + \delta \omega \partial_\omega \gamma_{\alpha \beta}(\omega) \left[ A_\beta(\omega) \rho A_\alpha^\dagger(\omega) - \frac{1}{2} \left\{ A_\alpha^\dagger(\omega) A_\beta(\omega), \rho \right\} \right].
\end{align} 
To analyse the size of these terms, we first estimate $\lambda^2\abs{ \gamma_{\alpha \beta}(\omega)} \sim \tau_R^{-1}$ for all $\omega$ of interest.
The first derivative can be found by observing that $\Gamma_{\alpha \beta}(\omega) = \int_0^\infty \dd s \; e^{i \omega s} \ev{B_\alpha^\dagger(s) B_\beta(0)}$ implies
\begin{align} \label{eqn:gamma_derivative}
    \abs{ \partial_\omega \Gamma_{\alpha \beta}(\omega) } & = \abs{ \int_0^\infty \dd s \; s e^{i \omega s} \ev{B_\alpha^\dagger(s) B_\beta(0)} }\nonumber \\
        & \sim \tau_B \abs{\Gamma_{\alpha \beta}(\omega)},
\end{align}
given that $\tau_B$ is the characteristic decay timescale of the correlation function~\cite{Albash2012Quantum}.
Therefore, we can estimate $\lambda^2 \abs{\partial_\omega \gamma_{\alpha \beta}(\omega)} \sim \tau_B \tau_R^{-1}$, and similarly for $S_{\alpha \beta}(\omega)$.
From the perturbation theory expressions above and the fact that $A_\alpha(\omega) = \mc{O}(1)$, we further have $A_\alpha^{(1)}(\omega) \sim h^{-1} \sim \tau_S$.
Hence, we see that
\begin{align} \label{eqn:weak_coupling_error2}
    v\mc{H}_\mathrm{LS}^{(1)}(\rho) + v \mc{D}^{(1)}(\rho) & = \mc{O} \left( \frac{\tau_S}{\tau_V \tau_R} \right) + \mc{O}\left( \frac{\tau_B}{\tau_V \tau_R} \right),
\end{align}
We want to ensure that this term is negligible compared with the other terms in the master equation.
It is small compared with $v$ as long as $\tau_S \ll \tau_R$ and $\tau_B \ll \tau_R$ -- exactly the RWA and Born-Markov conditions, assumptions (ii) and (i) above.
Furthermore, it is small compared with $\mc{D}$ and $H_\mathrm{LS}$ when $\tau_S \ll \tau_V$ and $\tau_B \ll \tau_V$, corresponding to the small perturbation assumptions (iv) and (iii).\\

Finally, we comment on what happens when a degeneracy in the Bohr frequencies is broken by the perturbation.
(The simplest example is a three-level system with energies $E_0=0,\, E_1= \varepsilon$ and $E_2 = 2\varepsilon$, with a perturbation shifting the middle level $E_1$.)
Suppose that $\abs{\omega_1 - \omega_2} \ll \tau_R^{-1}$ but $\abs{\omega'_1 - \omega'_2} \gg \tau_R^{-1}$.
This means the RWA now gets applied to remove the relevant off-diagonal term in the perturbed master equation.
Then the error term contains a part of order $\tau_R^{-1}$ -- of the same magnitude as $\mc{D}$ and thus non-negligible.
In order to prevent this from happening, in such a situation we therefore need the additional condition $\tau_R \ll \tau_V$ -- i.e., the perturbation must be weaker than the decoherence term.
This is stricter than the small perturbation conditions (iii) and (iv) assumed above.

\section{Error bounds for weak coupling} \label{app:errors}
\noindent \textbf{Result \ref{res:weak_coupling} (main text).}
\textit{
    For an open system in the weak coupling regime perturbed by the Hamiltonian $vV$,
    \begin{align} \label{eqn:weak_coupling_speed_app}
        \bures(\rho_t, \eta_t) \leq \frac{1}{2} \int_0^t \dd s \; v \sqrt{\mc{F}(\eta_s, \mc{V})} + \Delta(t),
    \end{align}
    where the error term is bounded by the estimate
    \begin{align} \label{eqn:weak_coupling_error_app}
        \abs{\Delta(t)} \lesssim \Delta_\mathrm{est}(t) := \frac{4\sqrt{2}}{3} \|V\| \epsilon^\frac{1}{2} (vt)^\frac{3}{2} + \epsilon v t.
    \end{align}
}

\begin{proof}
The triangle inequality for $\bures$ and an application of Result~\ref{res:main} (main text) implies
\begin{align}
    \bures(\rho_t, \eta_t) & \leq \bures(\rho_t, \sigma_t) + \bures(\sigma_t, \eta_t) \nonumber \\
        & \leq \frac{1}{2} \int_0^t \dd s \; v \sqrt{\mc{F}(\sigma_s,\mc{V})} + \bures(\sigma_t,\eta_t) \nonumber \\
        & \leq \frac{1}{2} \int_0^t \dd s \; v \sqrt{\mc{F}(\eta_s, \mc{V})} + \Delta_1(t) + \Delta_2(t),
\end{align}
where the error terms are
\begin{align} \label{eqn:two_deltas}
    \Delta_1(t) & := \frac{1}{2} \abs{ \int_0^t \dd s \; v \sqrt{\mc{F}(\sigma_s,\mc{V})} - \int_0^t \dd s \; v \sqrt{\mc{F}(\eta_s,\mc{V})} }, \nonumber \\
    \Delta_2(t) & := \bures(\sigma_t,\eta_t).
\end{align}
In order to bound these errors, we start by showing that $\Delta_1$ can be related to $\Delta_2$.
Firstly, using the triangle inequality and then the fact that $(\sqrt{x} - \sqrt{y})^2 = x + y - 2\sqrt{xy} \leq x + y - 2\min\{x,y\} = \abs{x-y}$,
\begin{align} \label{eqn:delta1_bound1}
    \Delta_1(t) & \leq \frac{v}{2} \int_0^t \dd s \; \abs{ \sqrt{\mc{F}(\sigma_s,\mc{V})} - \sqrt{\mc{F}(\eta_s,\mc{V})} } \nonumber \\
        & \leq \frac{v}{2} \int_0^t \dd s \; \sqrt{\abs{ \mc{F}(\sigma_s,\mc{V}) - \mc{F}(\eta_s,\mc{V}) }}.
\end{align}
Now we use the QFI continuity result Eq.~(A6) from Ref.~\cite{Augusiak2016Asymptotic}:
\begin{align}
    \abs{\mc{F}(\sigma,\mc{V}) - \mc{F}(\eta,\mc{V})} \leq 32 D_B(\sigma,\eta) \|V\|^2,
\end{align}
in terms of the Bures distance $D_B = \sqrt{2(1-F)}$ which satisfies $D_B \leq \bures$ (where $F$ is the Uhlmann fidelity).
Inserting this into Eq.~\eqref{eqn:delta1_bound1} gives
\begin{align} \label{eqn:delta1_bound2}
    \Delta_1(t) & \leq 2\sqrt{2} v \|V\| \int_0^t \dd s \; \sqrt{\Delta_2(s)}.
\end{align}
Now we bound $\Delta_2$ using another application of the speed limit in Result~\ref{res:main} (main text), now comparing $\sigma_s$ and $\eta_s$ under the error term $\mc{H}_\mathrm{LS}^{(1)} + \mc{D}^{(1)}$.
To lowest order in $v$,
\begin{align}
    \Delta_2(t) & \leq \frac{v}{2} \int_0^t \dd s \; \sqrt{\mc{F}(\eta_s, \mc{H}_\mathrm{LS}^{(1)} + \mc{D}^{(1)})}.
\end{align}
The size of the integrand is bounded using Lemma~\ref{lem:max_qfi} below, giving
\begin{align} \label{eqn:delta2_bound}
    \Delta_2(t) & \leq v t \, \max_\psi \|\mc{H}^{(1)}_\mathrm{LS}(\psi) + \mc{D}^{(1)}(\psi) \| \nonumber \\
        & =: v t \epsilon.
\end{align}
The right-hand side of Eq.~\eqref{eqn:delta2_bound} measures the size of the generator $\mc{H}_\mathrm{LS}^{(1)} + \mc{D}^{(1)}$ in terms of its greatest effect on any pure state.
In specific examples (see Appendix~\ref{app:qubits}) this norm can be calculated explicitly; from Eq.~\eqref{eqn:weak_coupling_error2}, we can generally estimate
\begin{align}
    \epsilon = \mc{O}\left(\frac{\tau_S}{\tau_R}\right) + \mc{O}\left( \frac{\tau_B}{\tau_R} \right).
\end{align}
Inserting into Eq.~\eqref{eqn:delta1_bound2}, we have
\begin{align}
    \Delta_1(t) & \leq 2\sqrt{2}v \|V\| \int_0^t \dd s \; \sqrt{v \epsilon s} \nonumber \\
        & = \frac{4\sqrt{2}}{3}\|V\| \epsilon^\frac{1}{2} (v t)^\frac{3}{2}.
\end{align}

\end{proof}

The validity and tightness of these estimates are shown in Fig.~\ref{fig:errors} for the two-qubit model described in Appendix~\ref{app:qubits}.

\begin{figure}[h]
    \includegraphics[width=0.55\textwidth]{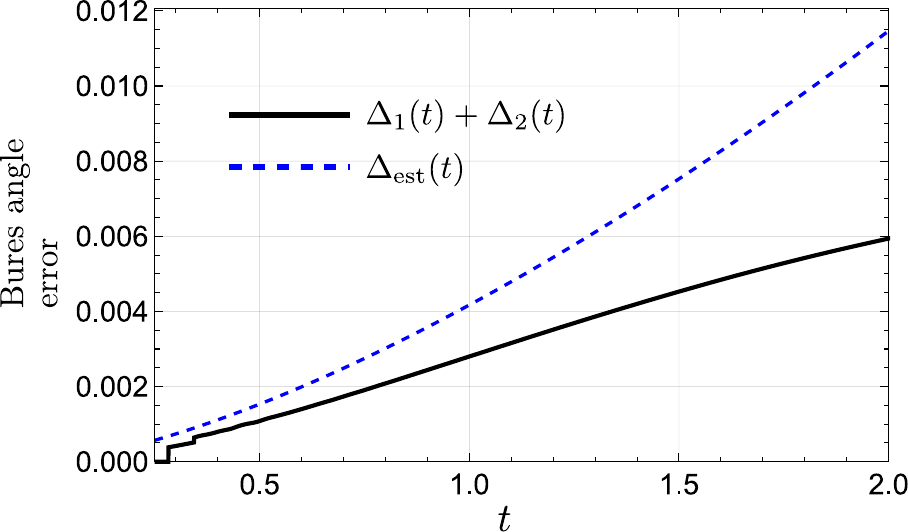}
    \caption{Demonstration of error bound for the two-qubit example, taking the same parameters as in Fig.~\ref{fig:qubits}.
    The error being shown is that on the right-hand side of the weak coupling speed limit Eq.~\eqref{eqn:weak_coupling_speed}: both the sum of the terms $\Delta_1(t) + \Delta_2(t)$ from Eq.~\eqref{eqn:two_deltas} (which in this case is dominated by $\Delta_2$) and the upper bound estimate $\Delta_\mathrm{est}(t)$ from Eqs.~\eqref{eqn:delta1_bound2}, \eqref{eqn:delta2_bound}.
    }
    \label{fig:errors}
\end{figure}

\begin{lem} \label{lem:max_qfi}
    For any state $\rho$ and generator $\mc{L}$, we have $\mc{F}(\rho, \mc{L}) \leq 4 \max_{\psi} \| \mc{L}(\psi) \|^2$, with the maximisation being over pure states $\psi$.
    \begin{proof}
        Convexity of QFI says that, for any pure state ensemble decomposition $\rho = \sum_i p_i \dyad{\psi_i}$, $\mc{F}(\rho,\mc{L}) \leq \sum_i p_i \mc{F}(\psi_i,\mc{L})$.
        Therefore the maximal QFI can always be achieved with some pure state $\psi$.
        We employ the expression $\mc{F}(\rho, \mc{L}) = 2 \sum_{i,j:\; \lambda_i + \lambda_j >0} \frac{\abs{ \mel{i}{\mc{L}(\rho)}{j} }^2}{\lambda_i + \lambda_j}$ with an orthogonal basis $\{\ket{i}\}_i$ chosen such that $\ket{0} = \ket{\psi}$, getting
        \begin{align}
            \mc{F}(\psi,\mc{L}) & = 2 \sum_{i,j} \frac{\abs{ \mel{i}{\mc{L}(\psi)}{j}}^2}{\lambda_i + \lambda_j} \nonumber \\
                & = 2 \left[ \frac{ \abs{\ev{\mc{L}(\psi)}{\psi}}^2}{2} + 2 \sum_{i>0} \abs{\mel{i}{\mc{L}(\psi)}{\psi}}^2 \right] \nonumber \\
                & = \ev{\mc{L}(\psi)}{\psi}^2 + 4 \sum_{i>0} \mel{\psi}{\mc{L}(\psi)}{i} \mel{i}{\mc{L}(\psi)}{\psi} \nonumber \\
                & = \ev{\mc{L}(\psi)}{\psi}^2 + 4 \bra{\psi}\mc{L}(\psi) \left(\id - \dyad{\psi} \right) \mc{L}(\psi) \ket{\psi} \nonumber \\
                & = 4 \ev{\mc{L}(\psi)^2}{\psi} - 3 \ev{\mc{L}(\psi)}{\psi}^2 \nonumber \\
                & \leq 4 \ev{\mc{L}(\psi)^2}{\psi}.
        \end{align}
        Finally, we note that $\ev{\mc{L}(\psi)^2}{\psi} \leq \|\mc{L}(\psi)^2\| \leq \|\mc{L}(\psi)\|^2$.
    \end{proof}
\end{lem}

\section{Two-qubit dephasing model} \label{app:qubits}
Our example is a two-qubit system in which each qubit dephases by interacting independently with a bath of harmonic oscillators -- see, for example, Ref.~\cite{lidar2019lecture}.
First consider a single qubit with Hamiltonian $H = \frac{h}{2} \sigma_z$ and perturbed with $V = \frac{1}{2} \sigma_x$, and a bath whose Hamiltonian is $H_B = \sum_k \omega_k b_k^\dagger b_k$, where $b_k$ is the bosonic annihilation operator for mode $k$.
The interaction is $H_I = \lambda \sigma_z \ox \sum_k g_k (b_k + b_k^\dagger)$, with $g_k$ being dimensionless coupling coefficients.
We take the continuum limit with an Ohmic bath, replacing $g_k^2$ by a spectral density function $J(\omega) = \eta \omega e^{-\omega/\omega_c}$.
Here, $\eta$ is a constant with dimensions of time squared and $\omega_c$ is a cut-off frequency that we take to be large compared with all other relevant frequency scales.
Given that the bath is in a thermal state at inverse temperature $\beta$, one can derive~\cite{lidar2019lecture}
\begin{align}
    \gamma(\omega) = \frac{2\pi \eta e^{-\abs{\omega}/\omega_c}}{1 - e^{-\beta \omega}}.
\end{align}
Since there is a single jump operator in the master equation, $A = A(0) = \sigma_z$, from the results in Appendix~\ref{app:weak}, we find $A^{(1)}(0) = \sigma_x$ and thus
\begin{align}
    H_\mathrm{LS}^{(1)} & = 0, \nonumber \\
    \mc{D}^{(1)}(\rho) & = \frac{\lambda^2 \gamma(0)}{h} \left( \sigma_z \rho \sigma_x + \sigma_x \rho \sigma_z \right).
\end{align}
Then $\gamma(0) = 2\pi \eta/\beta$ and $\tau_B \sim \beta$.
In this case, note that the derivative $\partial_\omega \gamma(\omega)$ does not appear.
However, for large $\omega_c$ one finds $\lim_{\omega\to 0^+}\partial_\omega \gamma(\omega) \approx \lim_{\omega\to 0^-}\partial_\omega \gamma(\omega) \approx \pi \eta \sim \gamma(0) \tau_B$ -- consistent with the claim in Eq.~\eqref{eqn:gamma_derivative}.
The error terms are easily bounded using
\begin{align}
    \| \sigma_z \psi \sigma_x + \sigma_x \psi \sigma_z\| & \leq 2 \| \sigma_z \psi \sigma_x \| \nonumber \\
        & \leq 2\| \sigma_z \| \| \psi \| \| \sigma_x\| \nonumber \\
        & = 2.
\end{align}
The single-qubit error parameter $\epsilon_1$ is therefore bounded by $\epsilon_1 \leq 2 \lambda^2 \gamma(0)/h$.\\

For the two-qubit system, we take the same independent dynamics on each qubit (so that $H = \frac{h}{2}(\sigma_z \ox \id + \id \ox \sigma_z)$ and so on).
This gives $\epsilon = \max_{\psi} \| [\mc{D}^{(1)} \ox \id + \id \ox \mc{D}^{(1)}](\psi)\| \leq 2 \epsilon_1 \leq 4 \lambda^2 \gamma(0)/h$.
Note that an extension of this model to $N$ qubits would have an error term scaling with $N$; a similar analysis could also be performed for a collective decoherence model where all qubits couple to the same bath~\cite{Kempe2001Theory}.

\section{Details of work fluctuations} \label{app:work}

Here, we first recall the derivation in Ref.~\cite{Miller2019Work} of Eq.~\eqref{eqn:fdr}.
It starts from writing the dissipated work in terms of relative entropy as $\beta W_\mathrm{diss} = S(\gibbs|| \gibbs') = \tr[(\gibbs - \gibbs') \ln \gibbs']$.
For small $\Delta H$, this can be approximated to lowest order as
\begin{align}
    W_\mathrm{diss} \approx \frac{\beta}{2} \variance^{\mathrm{K}}(\gibbs, \Delta H),
\end{align}
where $\variance^{\mathrm{K}}$ is the so-called Kubo-Mori generalised variance~\cite{Petz2002Covariance}.
This can be expressed using a superoperator $\mathbb{J}_\rho$ which depends on a given state $\rho$:
\begin{align}
    \mathbb{J}_\rho(A) = \sum_{i,j} \frac{\lambda_i - \lambda_j}{\ln \lambda_i - \ln \lambda_j} \mel{\psi_i}{A}{\psi_j} \dyad{\psi_i}{\psi_j},
\end{align}
where $\lambda_i$ and $\ket{\psi_i}$ are the eigenvalues and eigenvectors of $\rho$.
Then we have $\variance^{\mathrm{K}}(\rho,A) := \tr\left[\bar{A} \mathbb{J}_\rho (\bar{A})\right]$, with $\bar{A} := A - \tr[\rho A] \id$.
The crucial property is the splitting of the variance of $\Delta H$ into two terms: $\variance(\gibbs, \Delta H) = \variance^{\mathrm{K}}(\gibbs, \Delta H) + \bar{I}(\gibbs, \Delta H)$, interpreted as classical and quantum components respectively (see also Ref.~\cite{Frerot2016Quantum}).\\

We also require the connection between $\bar{I}$ and the QFI proved in Lemma~\ref{lem:qfi_Ibar_ineq} below.
This makes use of the theory of generalised QFI quantities -- see Ref.~\cite{Petz2002Covariance} for details.
Every generalised ``skew information" $I^f$ corresponds to a function $f \colon \mathbb{R}^+ \to \mathbb{R}^+$ fulfilling the conditions $f(1)=1$, $x f(x^{-1}) = f(x)$, and being a matrix monotone.
If $f(0) \neq 0$, it is possible to normalise $I^f$ to be ``metric-adjusted"~\cite{Hansen2008Metric}, such that $I^f(\dyad{\psi},H) = \variance(\dyad{\psi},H)$ for all pure states.
Explicitly, we have
\begin{align}
    I^f(\rho, H) & = \frac{f(0)}{2}\sum_{i,j} \frac{(\lambda_i - \lambda_j)^2}{\lambda_j f(\lambda_i/\lambda_j)} \abs{\mel{i}{H}{j}}^2.
\end{align}
The case $f(x) = \frac{x+1}{2}$ is often denoted by $I^\mathrm{SLD}$ (standing for ``symmetric logarithmic derivative") and recovers the standard QFI under evolution generated by $H$: $4 I^\mathrm{SLD}(\rho,H) = \mc{F}(\rho,\mc{H})$.

\begin{lem} \label{lem:qfi_Ibar_ineq}
    For all states $\rho$ and observables $H$,
    \begin{align}
        4\bar{I}(\rho,H) \leq \mc{F}(\rho,\mc{H}) \leq 12 \bar{I}(\rho,H).
    \end{align}
    \begin{proof}
        We start from the definition $I^{(k)}(\rho,H) := \frac{1}{2} \tr\left([\rho^k,H][H,\rho^{1-k}]\right)$, such that $\bar{I} = \int_0^1 \dd k \; I^{(k)}$.
        We can rewrite $I^{(k)}$ in terms of the matrix elements $H_{ij} := \mel{\psi_i}{H}{\psi_j}$ in the eigenbasis of $\rho = \sum_i \lambda_i \dyad{\psi_i}$:
        \begin{align}
            I^{(k)}(\rho,H) & = \tr \left( \rho H^2 - \rho^k H \rho^{1-k} H \right) \nonumber \\
                & = \sum_{i,j} (\lambda_i - \lambda_i^k \lambda_j^{1-k}) H_{ij} H_{ji} \nonumber \\
                & = \sum_{i,j} \frac{1}{2} (\lambda_i + \lambda_j - \lambda_i^k \lambda_j^{1-k} - \lambda_i^{1-k} \lambda_j^k) \abs{H_{ij}}^2 \nonumber \\
                & = \sum_{i,j} \frac{1}{2} (\lambda_i^k - \lambda_j^k)(\lambda_i^{1-k} - \lambda_j^{1-k}) \abs{H_{ij}}^2 \nonumber \\
                & = \frac{f^{(k)}(0)}{2} \sum_{i,j} \frac{(\lambda_i - \lambda_j)^2}{\lambda_j f^{(k)}(\lambda_i/\lambda_j)} \abs{H_{ij}}^2,
        \end{align}
        where $f^{(k)}$ is the matrix monotone function corresponding to $I^{(k)}$.
        We must therefore have
        \begin{align}
            f^{(k)}(x) \propto \frac{(x-1)^2}{(x^k - 1)(x^{1-k} - 1)}.
        \end{align}
        In order to be normalised in the standard way, we require $f^{(k)}(1) = 1$; taking the limit $x\to 1$ in the above shows that
        \begin{align}
            f^{(k)}(x) = \frac{k(1-k) (x-1)^2}{(x^k-1)(x^{1-k}-1)}
        \end{align}
        and $f^{(k)}(0) = k(1-k)$.
        For all metric-adjusted skew informations we have the inequality $I^f \leq I^\mathrm{SLD} \leq \frac{I^f}{2f(0)}$~\cite{Gibilisco2009Inequalities}.
        It is immediate from its definition that $I^{(k)}$ is metric-adjusted.
        Thus, the lower bound on $\mc{F} = 4I^\mathrm{SLD}$ is immediate, and we also have
        \begin{align}
            I^{(k)}(\rho,H) & \geq 2f^{(k)}(0) I^\mathrm{SLD}(\rho,H) \nonumber \\
                & = \frac{f^{(k)}(0)}{2} \mc{F}(\rho,\mc{H}), \nonumber \\
                & = \frac{k(1-k)}{2} \mc{F}(\rho,\mc{H}).
        \end{align}
        Finally,
        \begin{align}
            \bar{I}(\rho,H) & \geq \int_0^1 \dd k \; \frac{k(1-k)}{2} \mc{F}(\rho,\mc{H}) \nonumber \\
                & = \frac{1}{12} \mc{F}(\rho,\mc{H}).
        \end{align}
    \end{proof}
\end{lem}

\noindent \textbf{Result \ref{res:work} (main text).}
\textit{
    Quantum work fluctuations are necessary for fast departure from equilibrium.
    For a system weakly coupled to a thermal environment, at all times $t>0$ following the quench $H \to H'$, the distance between the initial state $\gibbs$ and the system's state $\rho_t$ obeys
    \begin{align} \label{eqn:work_speed_app}
        \bures(\gibbs, \rho_t) \leq t \sqrt{3 \bar{I}(\gibbs, \Delta H)} + \Delta(t),
    \end{align}
    where $\Delta(t)$ is the weak coupling error term from Eq.~\eqref{eqn:weak_coupling_error}.
}
\begin{proof}
    We start from Result~\ref{res:weak_coupling} (main text) with initial state $\gibbs$, but instead reversing the roles of the perturbed and unperturbed trajectories.
    So now the unperturbed trajectory involves the Hamiltonian $H'$ and its dynamics are generated by $\mc{L'}$, while the perturbation is $vV = - \Delta H$.
    Doing it this way around, the QFI appearing on the right-hand side of the speed limit is $\mc{F}(\gibbs,\Delta \mc{H})$, which is constant in time since the perturbed state $\rho_t =\gibbs \, \forall \, t\geq 0$ is a steady state under $\mc{L}$.
    The relevant Bures angle is then $\bures(\rho'_t, \rho_t) = \bures(\rho'_t, \gibbs)$.
    Lemma~\ref{lem:qfi_Ibar_ineq} shows that the QFI and $\bar{I}$ never differ by more than a constant factor: $4\bar{I} \leq \mc{F} \leq 12 \bar{I}$.
    Finally using the the latter inequality, we have $\mc{F}(\gibbs,\Delta \mc{H}) \leq 12 \bar{I}(\gibbs,\Delta H)$.
    \endproof
\end{proof}

\begin{figure}[h]
    \includegraphics[width=0.28\textwidth]{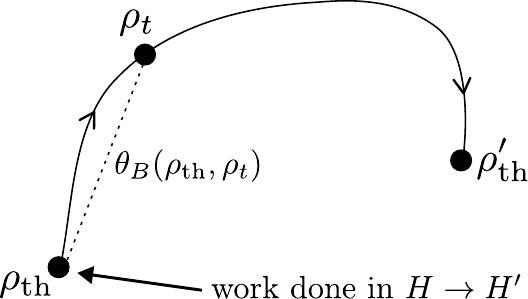}
    \caption{Driving out of equilibrium: work is performed during a sudden quench. The system moves away from its initial Gibbs state $\gibbs$ to the new one $\gibbs'$. Eq.~\eqref{eqn:work_speed_app} bounds the distance between $\gibbs$ and $\rho_t$ in terms of quantum work fluctuations.}
    \label{fig:work}
\end{figure}

\section{Bounds to linear response} \label{app:linear}

We take a system with time-independent unperturbed dynamics governed by $\mc{L}$, and a Hamiltonian perturbation $v_t V$ switched on for $t>0$.
At $t=0$, the system is assumed to be in a steady state $\pi$ of the unperturbed dynamics, so that $\mc{L}(\pi) = 0$.
Thus, as in Appendix~\ref{app:work}, the unperturbed trajectory is simply $\pi$ for all $t$, and we denote the perturbed trajectory by $\rho_t$.
For a given observable $A$, we are interested in the difference in its expectation value between the two trajectories,
\begin{align}
    \delta A_t := \tr[A(\rho_t - \pi)].
\end{align}
Firstly, from the main speed limit in Result~\ref{res:main} we can use the inequality $\abs{\tr[AB]} \leq \norm*{A} \norm*{B}_1$ to obtain
\begin{align}
    \abs{\delta A_t} & \leq \norm*{A} \norm*{\rho_t - \pi}_1 \nonumber \\
        & = 2 \norm*{A} D_{\tr}(\rho_t, \pi) \nonumber \\
        & \leq 2 \norm*{A} \bures(\rho_t, \pi) \nonumber \\
        & \leq \norm*{A} \int_0^t \dd s \, \sqrt{\mc{F}(\rho_s, \mc{P}_s)}.
\end{align}
Following the usual methodology of linear response~\cite{Kubo1957Statistical} we assume a small perturbation and can to lowest order approximate the QFI by its value in the unperturbed state $\pi$.
In addition, in a weak coupling setting, we can approximate $\mc{P}_s$ by $v_s \mc{V}$, with the error $\Delta(t)$ determined by Result~\ref{res:weak_coupling}:
\begin{align} \label{eqn:main_lim_response}
    \abs{\delta A_t} & \leq \norm*{A} \int_0^t \dd s \, \sqrt{\mc{F}(\pi,\mc{P}_s)} \nonumber \\
        & \lessapprox \norm*{A} \sqrt{\mc{F}(\pi,\mc{V})} \left[ \int_0^t \dd s\, \abs{v_s} + \Delta(t) \right].
\end{align}

Instead employing the observable speed limit in Result~\ref{res:obs_speed_lim}, we similarly have (for weak coupling and a weak perturbation)
\begin{align} \label{eqn:obs_lim_response}
    \abs{\delta A_t} & \leq \int_0^t \dd s \, 2\norm*{\mc{L}^\dagger(A)} D_{\tr}(\rho_s,\pi) + \sqrt{\variance(\rho_s,A) \mc{F}(\rho_s,\mc{P}_s)} \nonumber \\
        & \lessapprox \sqrt{\variance(\pi,A) \mc{F}(\pi,\mc{V})} \left[ \int_0^t \dd s \, \abs{v_s} + \Delta(t) \right],
\end{align}
having also used the fact that $D_{\tr}(\rho_s, \pi)$ is first-order in $s$ in order to drop the first drift term.
Although Eq.~\eqref{eqn:obs_lim_response} uses this additional approximation, it may be tighter than Eq.~\eqref{eqn:main_lim_response} due to the replacement of the norm $\norm*{A}$ by its fluctuations in the state $\pi$.
This is particularly true if $A$ has unbounded spectrum -- for instance, if it represents position or momentum of a particle -- in which case Eq.~\eqref{eqn:main_lim_response} becomes useless.
In general, we have
\begin{align}
    \variance(\pi,A) & \leq \tr[A^2 \pi] \leq \norm*{A^2} \norm*{\pi}_1 = \norm*{A}^2,
\end{align}
since $\norm*{\pi}_1 = 1$ for any quantum state $\pi$.

\end{document}